\documentclass[10pt,journal]{IEEEtran}
\usepackage{balance}
\usepackage[ruled]{algorithm2e}
\usepackage{amssymb}
\usepackage{cite}
\usepackage{color}
\usepackage{multirow}
\usepackage{graphicx,times,amsmath}
\usepackage{subfigure}

\usepackage{amssymb}
\usepackage{url}
\usepackage{makecell}
\usepackage{amsmath}

\ifCLASSOPTIONcompsoc
\else
\fi

\begin{document}
\title{Sensor-Assisted Rate Adaptation for UAV MU-MIMO Networks} 
\author{Xuedou~Xiao,~Wei~Wang,~\IEEEmembership{Senior Member,~IEEE,}~Tao~Jiang,~\IEEEmembership{Fellow,~IEEE}

\thanks{This work was supported in part by the National Key R\&D Program of China under Grant 2020YFB1806600, National Science Foundation of China with Grant 62071194, Tencent Rhino-Bird Focus Research Project of Basic Platform Technology 2021, and the European Union's Horizon 2020 research and innovation programme under the Marie Skłodowska-Curie grant agreement No 101022280. (\textit{Corresponding author: Wei Wang.)}}	

\thanks{X. Xiao, W. Wang, T. Jiang are with the School of Electronic Information and Communications, Huazhong University of Science and Technology, Wuhan 430074, China (e-mail: \{xuedouxiao, weiwangw, taojiang\}@hust.edu.cn).} }	
\maketitle

\begin{abstract}

Propelled by multi-user MIMO (MU-MIMO) technology, unmanned aerial vehicles (UAVs) as mobile hotspots have recently emerged as an attractive wireless communication paradigm. Rate adaptation (RA) becomes indispensable to enhance UAV communication robustness against UAV mobility-induced channel variances. However, existing MU-MIMO RA algorithms are mainly designed for ground communications with relatively stable channel coherence time, which incurs channel measurement staleness and sub-optimal rate selections when coping with highly dynamic air-to-ground links. In this paper, we propose SensRate, a new uplink MU-MIMO RA algorithm dedicated for low-altitude UAVs, which exploits inherent on-board sensors used for flight control with no extra cost. 
We propose a novel channel prediction algorithm that utilizes sensor-estimated flight states to assist channel direction prediction for each client and estimate inter-user interference for optimal rates. We provide an implementation of our design using a commercial UAV and show that it achieves an average throughput gain of 1.24$\times$ and 1.28$\times$ compared with the bestknown RA algorithm for 2- and 3-antenna APs, respectively.

\end{abstract}
	

\section{Introduction}\label{sec:intro}

\IEEEPARstart{O}{ne} key feature of the next generation communication system is the seamless cooperation between terrestrial and non-terrestrial infrastructures~\cite{kandeepan2014aerial}. Compared to terrestrial stations, unmanned aerial vehicles (UAVs), with their high flexibility and fast deployment, constitute a promising approach to provide on-demand, cost-effective and short-range wireless services. 
Numerous new UAV applications in civilian and commercial domains have emerged, such as traffic offloading in hotspot areas~\cite{chowdhery2018aerial,entrepreneur1,entrepreneur2,AT&T,bushnaq2020optimal}, delay-tolerant data collection from distributed wireless devices~\cite{samir2019uav,bushnaq2019aeronautical} and UAV-enabled mobile edge computing~\cite{zhang2018stochastic,zhou2018computation}, which normally require intensive uplink transmission. Driven by the need for superior spectral efficiency and network capacity, numerous efforts have been devoted to applying MU-MIMO technology to air-to-ground systems~\cite{8657707,9082100,8407088}.

Due to the high flexibility, low cost and easy accessibility, low-altitude small UAVs have always been favored by companies~\cite{AT&T,entrepreneur1,entrepreneur2}.	
We focus on the low-altitude small UAVs to provide short-range high-rate wireless service over a very small cell. 
The Federal Aviation Administration (FAA) guidelines may be used as a working definition for this category: UAVs can freely fly below the altitude of 120~m without any permit~\cite{FAA}. In this category, AT\&T~\cite{AT&T} and multiple companies~\cite{entrepreneur1,entrepreneur2} have tested the flying station at an altitude of 60~m. DroneFi~\cite{chowdhery2018aerial} proposes UAV hotspots flying below 20~m over a smaller cell size. The UAV altitudes set in the studies of UAV-enabled data collection and mobile edge computing generally vary from 10~m to 100~m~\cite{zhang2018stochastic,zhou2018computation,samir2019uav,bushnaq2019aeronautical}. Some researchers argue that the lower altitude, the less UAV energy consumption and more stable high-rate connection~\cite{zorbas2016optimal}.

In the context of low-altitude UAV hotspots, ensuring high-quality network services (i.e., high-rate, ultra-reliable) under dynamic flight states is paramount. Yet, air-to-ground links suffer from time-varying channels induced by agile UAV mobility and fast wireless fading. Current communication systems typically enhance communication robustness through rate adaptation (RA). However, exiting RA algorithms~\cite{6990336,xie2013adaptive,turborate,Scalable,sur2016practical,lin2017acpad} are not suitable for UAV hotspots with MU-MIMO networks to serve multiple clients. 
MU-MIMO RA algorithms~\cite{6990336,xie2013adaptive,turborate,Scalable,sur2016practical,lin2017acpad} are mainly for ground-to-ground communications, which target stationary devices, pedestrians, vehicles with relatively stable velocity and channel coherence time. They measure channel information in pre-estimated coherence time for optimal rates. 
However, the varying flight states cause the coherence time to continuously change~\cite{chowdhery2018aerial}, leading to severe channel measurement staleness and sub-optimal rate selections. 

A rich body of literature has been devoted to channel prediction in UAV scenarios. Whereas, they focus more on the received power, the amplitude and the throughput~\cite{chowdhery2018aerial,jiang2020learning,7835273,8048502,staterate,sa-abr}, but rarely involve the prediction of phase.
In MU-MIMO networks, there exists inter-user interference due to the non-orthogonality of clients' channel directions, which deeply affects network capacity and rate selections. 
The channel direction depends on both phase and amplitude. In particular, 
slight UAV movements can cause significant changes in phase. Willink et al.~\cite{7079507} take into account the phase by measuring the spatial correlation across the antenna array along the flight path, but do not form an effective prediction method.

In this paper, we propose SensRate, a new uplink MU-MIMO RA algorithm dedicated for low-altitude UAVs, which exploits UAV's inherent on-board sensors to predict channel directions and inter-user interference for optimal rates. The key observation is that the fluctuation patterns of channel directions are closely related to the UAV's movements. Through theoretical analysis and real-world measurements, we find that the phase difference of the antenna array that determines the channel direction significantly decreases or increases when wireless fading occurs along flight trajectories, while remaining relatively stable without fading. Contrary to the extreme sensitivity of phase, the phase difference exhibits more stable and regular changes, facilitating the prediction of the channel direction. 
Based on this observation, we first model the fine-grained changes in clients' channel directions along UAV's trajectories over time, from both phase and amplitude perspectives. 
Then, we propose a sensor-assisted channel prediction scheme that estimates the wireless fading frequency as a function of flight states and predicts the channel direction for each client. Finally, SensRate enables each client to estimate interference from others and pick accurate rates to maximize the overall system throughput.

We implement SensRate on a DJI Matrice 100 and compare our design with baseline algorithms under a wide range of conditions, including varying flight states, CSI reading rates, environments and moving clients. The experimental sites include an empty square and a parking lot with different reflectors nearby. 
The flight states of UAV cover velocities of 0-10~m/s and link distances ranging from 5-45~m with random locations of clients, reaching a total of 2.5-hour aerial CSI measurement. 
The results show that SensRate achieves an average throughput gain of 1.24$\times$ and 1.28$\times$ over the best-known RA algorithm for 2- and 3-antenna APs, respectively.

The contributions are summarized below.

\begin{itemize}

\item Through theoretical analysis and real-world experimental measurements, we thoroughly investigate the channel direction changing pattern over time as the UAV moves. Unlike previous works~\cite{chowdhery2018aerial,jiang2020learning,7835273,8048502,staterate,sa-abr} that generally focus on the received power, amplitude and throughput, we parameterize the change in channel directions, from both amplitude and phase perspectives, to support the prediction of inter-user interference in UAV MU-MIMO networks.

\item We propose a sensor-assisted channel prediction scheme that exploits the UAV's inherent sensor data to better adapt to the highly dynamic air-to-ground links. 
Unlike prior works that rely on coarse-grained distance regimes~\cite{8432501,8125764, sa-abr},
we estimate the wireless fading frequency/interval as a function of UAV velocities and positions, which improves the adaptability of SensRate by effectively mitigating the negative impacts caused by varying flight states.

\item We propose a new uplink MU-MIMO RA algorithm dedicated for UAVs to maximize the overall throughput. Unlike previous MU-MIMO RA algorithm that are susceptible to UAV-mobility induced channel variances~\cite{6990336,xie2013adaptive,turborate,Scalable,sur2016practical,lin2017acpad}, SensRate is built based on the sensor-assisted channel prediction scheme, which effectively mitigates the impact of CSI staleness caused by the fast wireless fading and agile UAV mobility.

\end{itemize}

The remainder of the paper is organized as follows. Section~\ref{sec:background} explores the impact of different flight states on MU-MIMO networks, which is the underpinning of our design. Section~\ref{sec:design} elaborates on the design of SensRate. Section~\ref{sec:evaluation} presents the implementation details and evaluation results. Section~\ref{sec:related} reviews the related works, followed by some discussions in Section~\ref{discussion} and the conclusion in Section~\ref{conclusion}.

\section{Exploring Flight Impact on MU-MIMO}\label{sec:background}

Recently, research efforts have been made toward deploying UAV hotspots to fly contiguously to shorten the link distances and avoid the blockage of buildings and trees~\cite{samir2019uav,zhang2018stochastic,chowdhery2018aerial}. 
Besides, the UAV mobility has been verified to consume less battery than UAV hovering. Thus, we focus on the RA design in mobile UAV scenarios,  
yet the profound impact of varying flight states on MU-MIMO networks is still under-explored.

In this section, we first introduce the generation of inter-user interference in MU-MIMO networks. 
Then, we take a deep dive into the impact of various flight states on MU-MIMO networks, which prompts us to incorporate flight-state-related sensor data into SensRate design.

\subsection{Interference by Non-Orthogonal Channel Directions}

Suppose a scenario where two single-antenna ground clients concurrently communicate with a 2-antenna UAV.
The received signals $\textbf{y} = (y_1, y_2)$ on the UAV is formulated as

\begin{equation}
\begin{pmatrix} y_1 \\  y_2\end{pmatrix} = \begin{pmatrix} h_{11} & h_{21} \\ h_{12}& h_{22}\end{pmatrix}\begin{pmatrix} x_1 \\  x_2\end{pmatrix}+\begin{pmatrix} n_1 \\  n_2\end{pmatrix},
\end{equation}
where $\textbf{h}_k = (h_{k1}, h_{k2})$ denotes the channel vector between client $k$ and AP, $n_m$ the noise at AP's antenna $m$, which follows $n_m\sim\mathcal C\mathcal N(0, N_0)$. 
The AP utilizes a zero-forcing (ZF) technique~\cite{turborate} to decode $x_1$, $x_2$ by projecting $\textbf{y}$ along a direction orthogonal to the channel vector of one client, say client 1. This allows AP to first decode $x_2$, but results in a signal to noise ratio (SNR) reduction caused by inter-user interference. According to \cite{turborate, signpost}, we can compute the resulting $\text{SNR}_{proj}$ and $\Delta\text{SNR}$ by
\begin{equation}\label{sinr}
\text{SNR}_{proj} = \text{SNR}_{orig}(1-{\cos^2\theta}),
\end{equation}
\begin{equation}\label{sinr2}
\Delta\text{SNR} (dB) = -10 \log_{10}(1-\cos^2\theta),
\end{equation}
where $\text{SNR}_{orig}$ is the original SNR of $x_2$ when the client transmits alone, i.e., without projection, $\theta$ is the angle between the directions of channel vectors $\textbf{h}_1$, $\textbf{h}_2$, that is

\begin{equation}\label{angle}
\cos^2\theta =\frac{|\textbf{h}_1\cdot \textbf{h}_2|^2}{\Vert\textbf{h}_1\Vert^2\Vert\textbf{h}_2\Vert^2}.
\end{equation}
We can see the detailed derivation process and experimental verification in \cite{turborate}. When extended to scenarios with an $M$-antenna UAV and $K (K \leq M)$ concurrent streams, $\theta$ becomes the angle between the channel direction of one client, say $\textbf{h}_k$ and the subspace $S$ spanned by other $K-1$ directions of concurrent streams. The channel direction of the client $k$ is defined as the direction of its channel vector $\textbf{h}_k$~\cite{turborate}. We can observe that once $\theta$ is less than $45^{\circ}$, $\Delta\text{SNR}$ exceeds 3~dB, and when $\theta$ is $10^{\circ}$, $\Delta \text{SNR}$ reaches 40~dB.
The inter-user interference deeply affects the optimal rates and throughput.

\begin{figure}
	\centering
	\subfigure[Impact of velocity]{
		\includegraphics[width=0.48\linewidth]{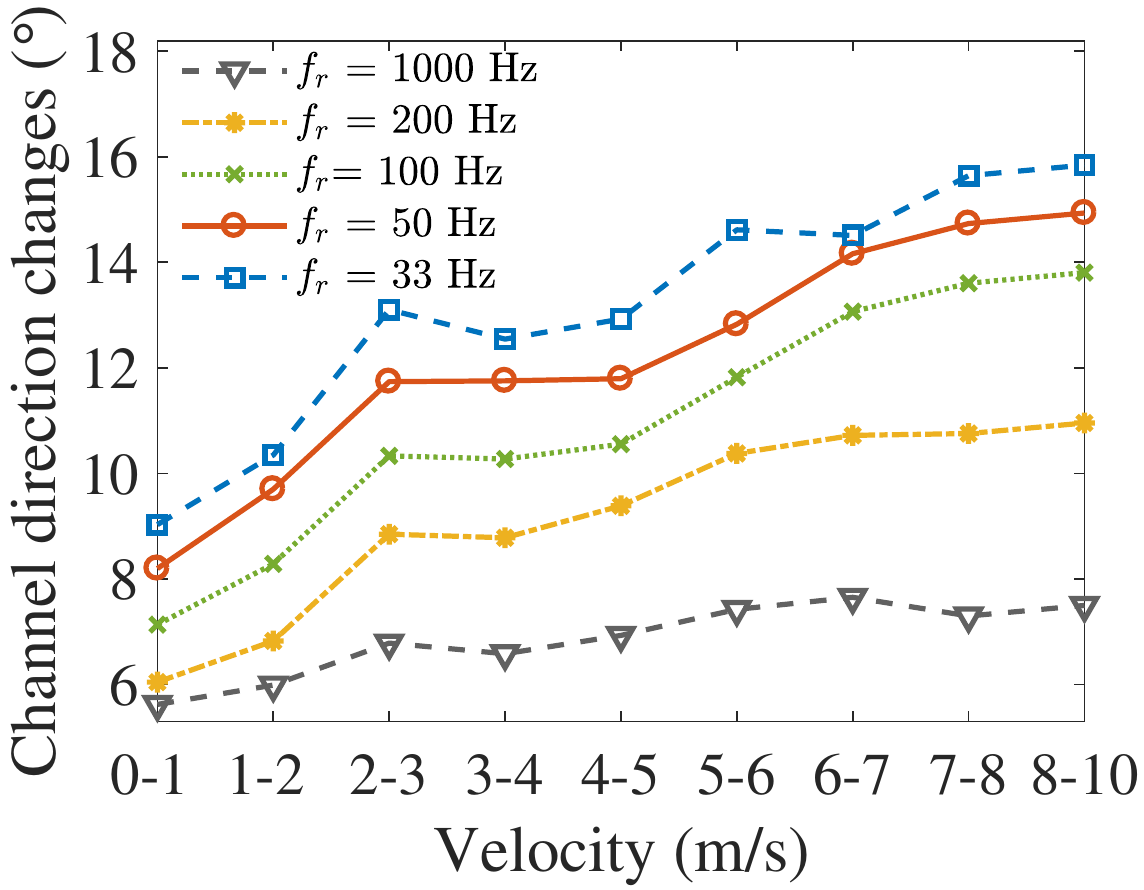}\label{section2-vel}}
	\hfill
	\subfigure[Impact of distance]{
		\includegraphics[width=0.48\linewidth]{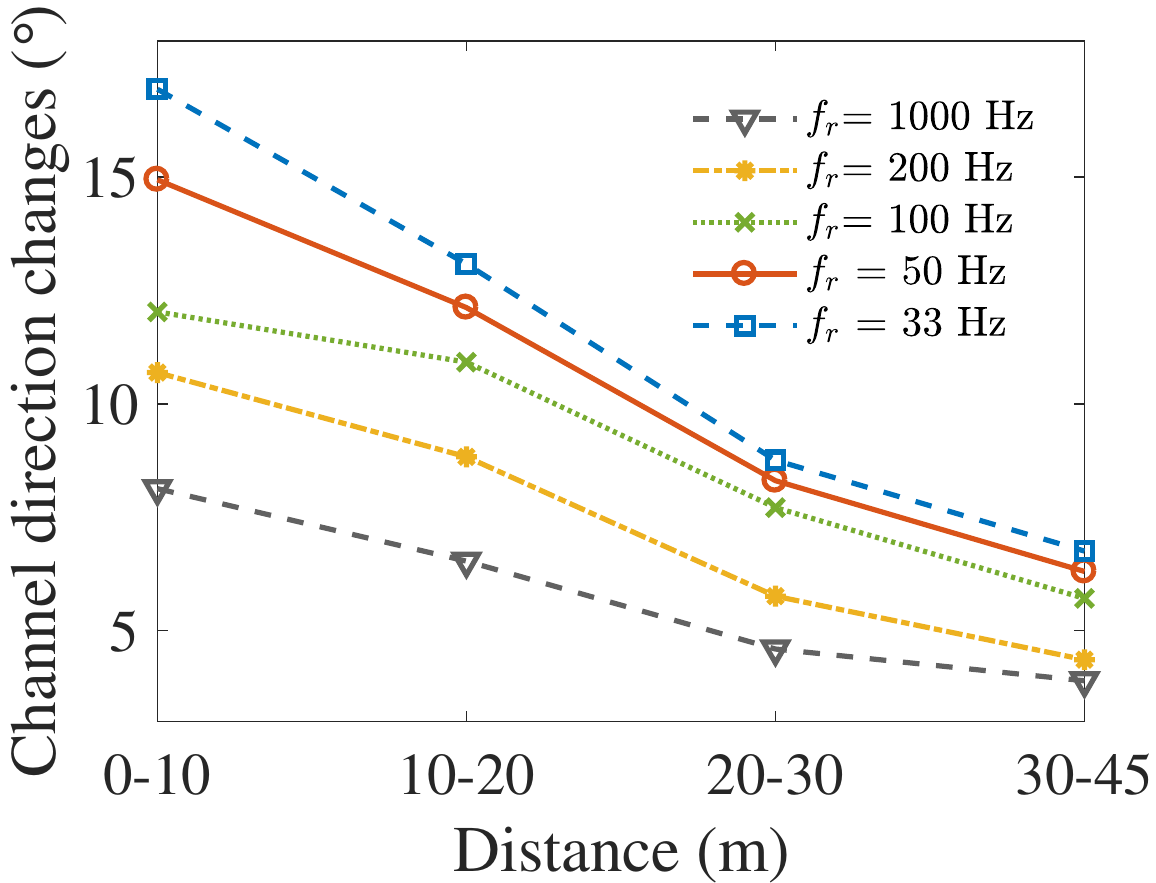}\label{section2-dis}}
	\\
	\vspace{-0.2cm}\caption{Impact of different flight states and CSI reading rates on channel direction changes.}\vspace{-0.3cm}
	\vspace{-0.1cm}\label{section2-vel-dis} 
\end{figure}

\subsection{Impact of UAV's Flight States}\label{sec:background.b}

Since the overall throughput is closely related to the inter-user interference, we conduct a series of experiments to investigate the impact of varying flight states on the channel direction changes.
We deploy a DJI Matrice 100 UAV as a multi-antenna AP to log sensor readings and 
collect the UAV-to-client CSI at different rates $f_r$ including 1000~Hz, 200Hz, 100~Hz, 50Hz and 33Hz. 
We measure the channel direction changes between adjacent CSI readings to represent the speed of channel variations under different flight states.

\textbf{Impact of velocity.} Fig.~\ref{section2-vel} plots 
the average channel direction changes during different CSI reading intervals at various UAV's velocities from 0~m/s to 10~m/s. The UAV-to-client distance is limited within 10-20~m. As shown in Fig.~\ref{section2-vel}, the channel direction changes during the same CSI reading interval increase with the UAV velocity. Once $f_r \leq $ 100~Hz, any UAV velocity above
2~m/s can cause a $10^{\circ}$ channel direction variation during the reading interval, and velocities more than 7 m/s may cause a variation of nearly $14^{\circ}$. 
According to Eq.~\eqref{sinr}-\eqref{angle}, $10^{\circ}$ channel direction variation during the CSI reading interval can render a change in $\text{SNR}_{proj}$ ranging from 0 to 20 dB, which deeply affects the optimal rate selections.

\textbf{Impact of distance.} Likewise, Fig.~\ref{section2-dis} plots the average channel direction changes during different CSI reading intervals under various UAV-to-client distance ranges. The UAV velocity is limited within 2-4~m/s. We observe that closer UAV-to-client distances are more likely to cause larger channel direction variations during each CSI reading interval.
When the distance is less than 30~m, average channel direction changes of $8^\circ$ are measured at $f_r \leq $ 100~Hz.

\begin{figure}
	\centering
	\subfigure[Channel direction changes]{
		\includegraphics[width=0.48\linewidth]{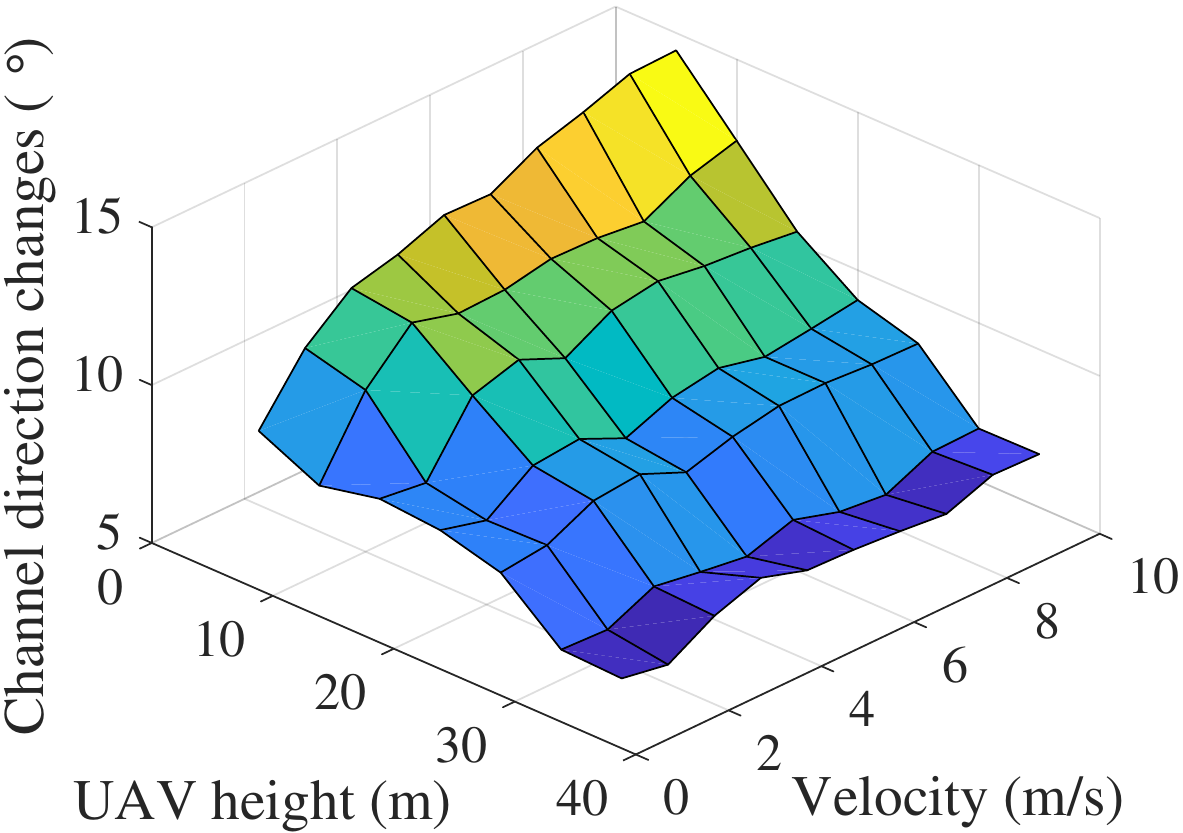}\label{section2-3d-direction}}
	\hfill
	\subfigure[SNR changes]{
		\includegraphics[width=0.48\linewidth]{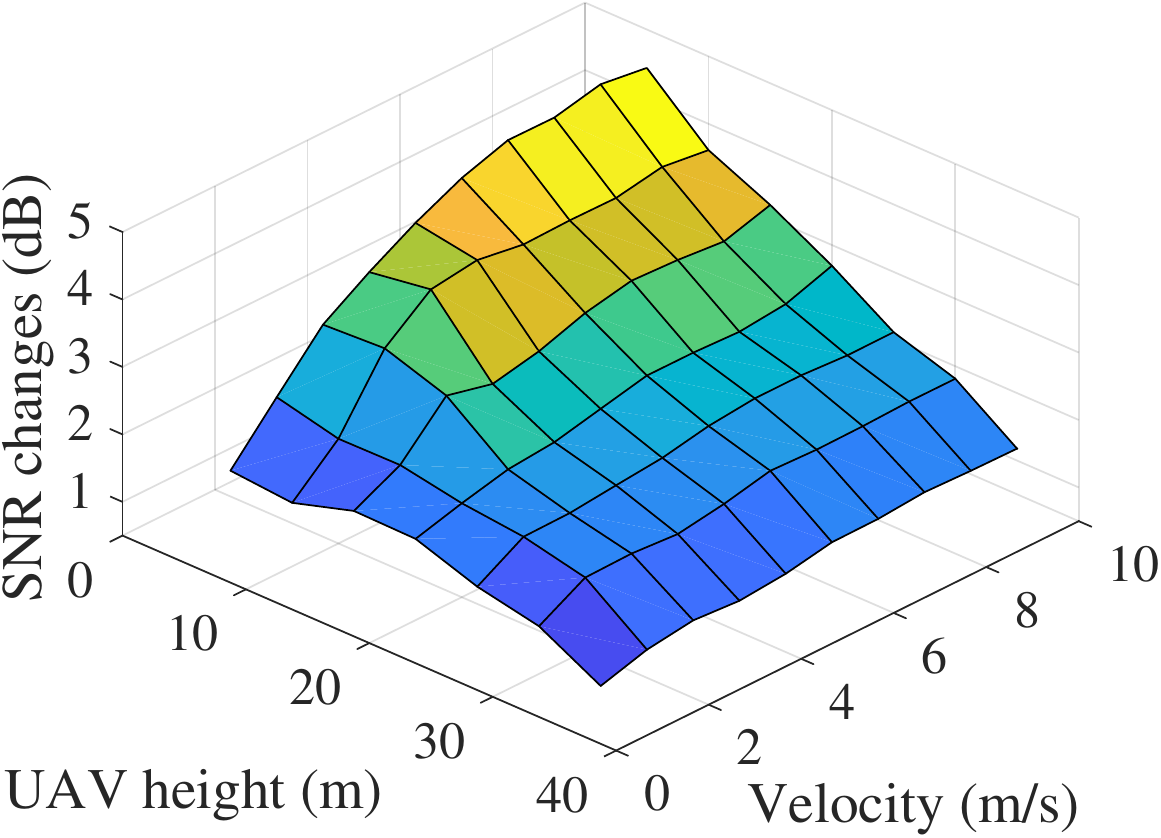}\label{section2-3d-SNR}}
	\\
	\vspace{-0.2cm}\caption{Impact of UAV velocity and height on channel changes during a CSI reading interval of 20~ms.}\vspace{-0.2cm}
	\vspace{-0.1cm}\label{section2-3d} 
\end{figure}

Fig.~\ref{section2-3d-direction} and Fig.~\ref{section2-3d-SNR} further demonstrate the combined impact of UAV velocity and altitude. 
The z axes respectively show the average changes in the channel direction and SNR during 20~ms ($f_r =$ 50~Hz). Note that the average change in channel direction or SNR during 20~ms decreases with UAV altitude, but increases with UAV velocity. 
At a low altitude of 10~m, a velocity of 2~m/s will result in an average channel direction change of 9$^\circ$ and an average SNR change of nearly 2~dB during 20~ms. However, as the UAV altitude increases to 40~m, the impact of UAV velocity becomes less significant. 
Only a much higher velocity of 10~m/s may cause a channel change during 20~ms comparative to that at low altitudes at a velocity of 1-2~m/s.
 
To summarize, the speed of channel variations changes rapidly with varying flight states. Once the CSI reading rates fail to adapt to the speed of channel variations, 
the traditional MU-MIMO RA algorithms that simply utilize the past channel information for rate selections fall short. Instead, we need a channel prediction scheme that
can leverage the flight-state-related sensor data to predict $SNR_{proj}$ for each client.

\section{UAV MU-MIMO Rate Adaptation}\label{sec:design}

In this section, we first characterize the overall SensRate architecture. 
Then, we model the channel direction dynamics as the UAV moves and propose a sensor-assisted prediction algorithm that predicts $\text{SNR}_{proj}$ over the next few milliseconds.
Based on the predicted $\text{SNR}_{proj}$, we proceed to execute the RA scheme and choose the accurate rates for all concurrent MU-MIMO clients on a per-packet basis to maximize the overall throughput of the uplink (UL) transmission.

\subsection{System Architecture}\label{sec:design.a}

\begin{figure}[t]
	\centering
	\includegraphics[width=0.48\textwidth]{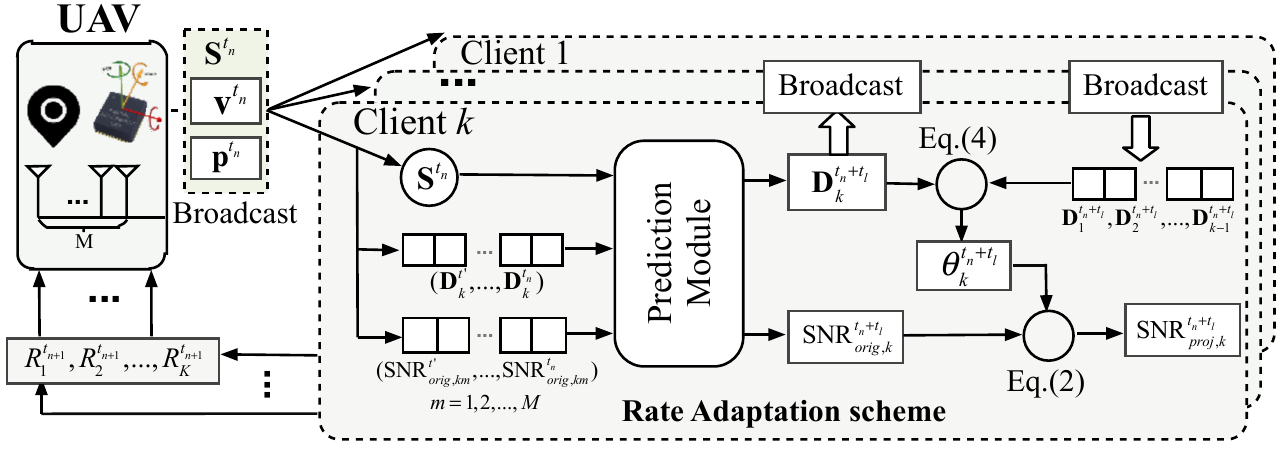}
	\caption{Overall SensRate architecture.}\label{model}\vspace{-1mm}
\end{figure}

\begin{figure}[t]
	\centering
	\includegraphics[width=0.475\textwidth]{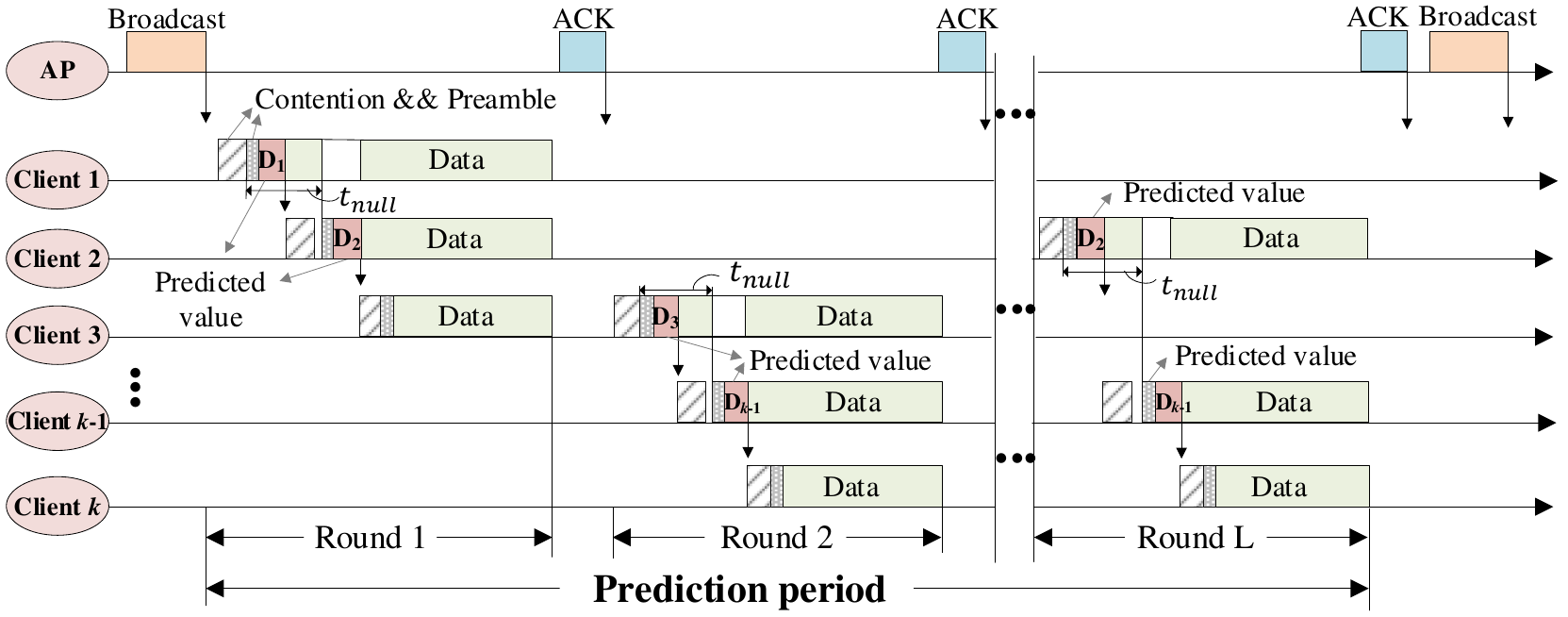}
	\caption{The prediction period in SensRate.}\label{prediction_time}\vspace{-3mm}
\end{figure}

Fig.~\ref{model} shows the overall architecture of SensRate at a high level. We consider the scenario where $K(K\leq M)$ single-antenna ground clients concurrently communicate with an $M$-antenna UAV hotspot. 
Note that each ground client $k(k\in\{1,2,...,K\})$ keeps listening to the periodic sensor broadcast 
from the UAV. They obtain the real-time flight states $\textbf{S}$ of the UAV and passively learn their UL channel vectors $\textbf{h}_k=(h_{k1},h_{k2},...,h_{kM})$ using channel reciprocity. Therein, we simplify the direction of UL channel as $\textbf{D}_k = (1, h_{k2}/h_{k1},...,h_{kM}/h_{k1})$ and each $\text{SNR}_{orig, km}$ can be computed by $|h_{km}|^2 P/N_0$, where $P$ is the client's transmission power, and $N_0$ the average noise level at the AP. 
Then, we input the past measurements of $(\text{SNR}_{orig,km}^{t'},...,\text{SNR}_{orig,km}^{t_n})$ $(\textbf{D}_k^{t'},...,\textbf{D}_k^{t_n})$ and the sensor data $\textbf{S}^{t_n}$ into the channel prediction module, where $t', t_n$ refers to the time of broadcast from $t'$ to $t_n$ and $m=1,2,...,M$.
The prediction module separately predicts the changes in $\text{SNR}_{orig,k}$, $\textbf{D}_k$ for each client $k$ in the following rounds of transmission before the next CSI measurement is available, as illustrated in Fig.~\ref{prediction_time}. The sensor broadcast rate here indicates the minimum rate of getting CSI readings for ground clients.

Next, similar to the MAC protocol in SAM~\cite{Tan2009SAM}, clients join concurrent transmissions one after another. They count the number of concurrent streams by cross-correlating with the known preamble to detect whether the number of existing streams equals $M$. 
The client that wins the transmission opportunity 
announce the predicted directions $\textbf{D}^{t_{n}+t_{l}}$ to later contenders by annotating the physical layer convergence protocol (PLCP) header\footnote{According to \cite{turborate}, in order to enable later contenders to decode the direction information, the ongoing transmissions will pause their streams at predefined times $k*t_{null}\ (k = 1,2,...,M-2)$ and send null samples for a period of time that is long enough to broadcast the direction information.}, according to~\cite{turborate}. Therein, $t_{n}+t_{l}$ represents the time for the round $l$ of the transmission and $t_n+t_l<t_{n+1}$. The clients that can hear this channel direction broadcast are eligible to continue to join the contention. Otherwise, they will give up the contention and wait for the next round of transmission. 
Then, the later winners can learn the predicted channel directions of already ongoing streams and combine its own channel predictions to calculate the angle $\theta_k^{t_{n}+t_{l}}$ 
and resulting $\text{SNR}_{proj,k}^{t_{n}+t_{l}}$ by Eq.~\eqref{sinr}-\eqref{angle}. 
Depending on $\text{SNR}_{proj,k}^{t_{n}+t_{l}}$, the rate $R_k^{t_{n}+t_{l}}$ is determined for each client to maximize the overall throughput, even if the clients change in different round of transmission. Note that the clients contend for the medium using the traditional 802.11 content mechanism in our experiment. However, more advanced contention and user selection mechanisms like \cite{mimomate,signpost} can also be adopted in SensRate, but this is beyond the scope of our work.

We formulate the maximization of overall throughput as  
\begin{equation}\label{overall_throughput}
\max_{R_1,\cdots, R_K} \sum^K\eta_k\left[R_k, p_k(\text{SNR}_{proj,k}, R_k)\right],
\end{equation}
\begin{equation}\label{sinrk}
\text{SNR}_{proj,k}= f(\textbf{h}_k^{t'},...,\textbf{h}_k^{t_n},\textbf{D}_1^{t_{n}+t_{l}},...,\textbf{D}_{k-1}^{t_{n}+t_{l}},\textbf{S}^{t_n}),
\end{equation}
where throughput $\eta_k$ for each client $k$ depends on the selected rate $R_k$ and bit error rate (BER) $p_k$, and $p_k$ is further determined by $\text{SNR}_{proj,k}$ and $R_k$. Therein, the predicted $\text{SNR}_{proj,k}$ at the time $t_{n}+t_{l}$ is decided by the past channel measurements $(\textbf{h}_k^{t'},...,\textbf{h}_k^{t_n})$ from time $t'$ to $t_n$, the predicted channel directions of the already ongoing clients ($\textbf{D}_1^{t_{n}+t_{l}},...,\textbf{D}_{k-1}^{t_{n}+t_{l}}$) and the sensor data $\textbf{S}^{t_n}$, as described above.

\subsection{Channel Direction Modeling}\label{sec:design.b}
Before proposing the sensor-assisted prediction algorithm, we first explore how the channel direction $\textbf{D}$ changes with UAV movements in this section. 

\textbf{Near regime.} 
Mobile UAV hotspots can leverage flying capabilities to provide short-range and unobstructed wireless network.
The resulting transmission regime overlaps with the Fresnel zone~\cite{chowdhery2018aerial,312809}, where
the path loss is dominated by the constructive and destructive interference between the line-of-sight path and propagation paths from nearby reflectors.
Thus, we focus on the $\textbf{D}_k$ changes under multipath effect.
Recall that $\textbf{D}_k$ is defined as $(1, h_{k2}/h_{k1},...,h_{kM}/h_{k1})$
, where each $h_{km}/h_{k1} (m\in\{1,2,...,M\})$  can be written in the form of amplitude and phase
\begin{equation}\label{phase_minus}
h_{km}/h_{k1} = |h_{km}|/|h_{k1}|e^{j(\varphi_{km}-\varphi_{k1})}.
\end{equation}
Therein, $\Delta\varphi_{km} =\varphi_{km}-\varphi_{k1}$ denotes the phase difference between signals received by antenna $1$ and $m$ on the UAV from client $k$. We convert $\textbf{D}_k$ to separate analyses of $\Delta\varphi_{km}, |h_{km}|$ and focus on the change in $\Delta\varphi_{km}$ in this subsection. The prediction of $|h_{km}|$ is detailed in the next subsection.

\textbf{Two-ray ground propagation model.}
Our analysis of the multipath effect starts with a two-ray ground propagation model. 
Suppose that the horizontal distance between the UAV and the client is $d_{H}$ and the heights of the UAV and the client are $d_U$ and $d_{c}$, respectively. Then, each $h_{km}$ between the client $k$ and the antenna $m$ on the UAV can be modeled by
\begin{equation}\label{two-ray-eq}
\begin{split}
h_{km} &=  h_d+h_r \\
&=\frac{1}{d_d}\exp(-\frac{j2\pi d_d}{\lambda}) + \rho\frac{1}{d_r}\exp(-\frac{j2\pi d_r}{\lambda}),
\end{split}
\end{equation}
where $h_d$ denotes the direct-path channel, $h_r$ the reflected-path channel, $\rho$ the reflection coefficient from the ground\footnote{The reflection coefficient from the ground $\rho$ is within -1 to 0, and is close to -1 when the ground surface is paved and asphalt.}, $d_d$ and $d_r$ the link distances of the direct path and the reflected path. 
We assume $d_r= \gamma d_d$ and the phase $\varphi_{km}$ can be extracted as
\begin{equation}\label{phase_2ray}
\varphi_{km} =  -\frac{2\pi d_d}{\lambda}-\arctan{\frac{\frac{\rho}{\gamma}\sin{\frac{2\pi(\gamma-1)d_d}{\lambda}}}{1+\frac{\rho}{\gamma}\cos{\frac{2\pi(\gamma-1)d_d}{\lambda}}}}.
\end{equation}
Then, we derivate Eq.~\eqref{phase_2ray} with respect to $d_d$ as follows
\begin{equation}
\varphi_{km}'|_{d_{d}}=-\frac{2\pi}{\lambda}-\frac{\pi}{\lambda}a_0-\frac{-\frac{\rho\gamma'}{\gamma^2}\sin{\frac{2\pi(\gamma-1)d_d}{\lambda}+\frac{\pi}{\lambda}(\frac{\rho^2}{\gamma^2}-1)a_0}}{1+\frac{\rho^2}{\gamma^2}+\frac{2\rho}{\gamma}\cos{\frac{2\pi(\gamma-1)d_d}{\lambda}}},
\end{equation}
where $a_0=\gamma'd_d+\gamma-1$. When $h_d$ and $h_r$ interfere destructively,  $\varphi_d$ and $\varphi_r$ gradually reach the state: $\varphi_d \approx \varphi_r + (2\beta+1)\pi, \beta\in \mathbb{Z} $. Therefore, we calculate $d_d$ during a fading period by $d_d=\frac{\beta\lambda}{\gamma-1}$. Similarly, when $h_d$ and $h_r$ interfere constructively, $\varphi_d$ and $\varphi_r$ gradually reach the state: $\varphi_d \approx \varphi_r + 2\beta\pi, \beta\in \mathbb{Z} $.  Thus, $d_d = \frac{(2\beta+1)\lambda}{2(\gamma-1)}$. As a result, $\varphi_{km}'|_{d_{d}}$ in this two states can be simplified to
\begin{equation}
\varphi_{km}'|_{d_d}=
\begin{cases}
	-\frac{2\pi}{\lambda}(1+a_0\frac{1}{1+\frac{\gamma}{\rho}}), & d_d=\frac{\beta\lambda}{\gamma-1}.\\
	-\frac{2\pi}{\lambda}(1+a_0\frac{1}{1-\frac{\gamma}{\rho}}),& d_d= \frac{(2\beta+1)\lambda}{2(\gamma-1)}.
\end{cases}
\end{equation}
Then we analyze the change in $\varphi_{km}'|_{d_{d}}$ from two different cases when the UAV flies in the horizontal or vertical direction. 

\begin{figure}[t]
	\centering
	\subfigure[Horizontal flight]{
		\includegraphics[width=0.48\linewidth]{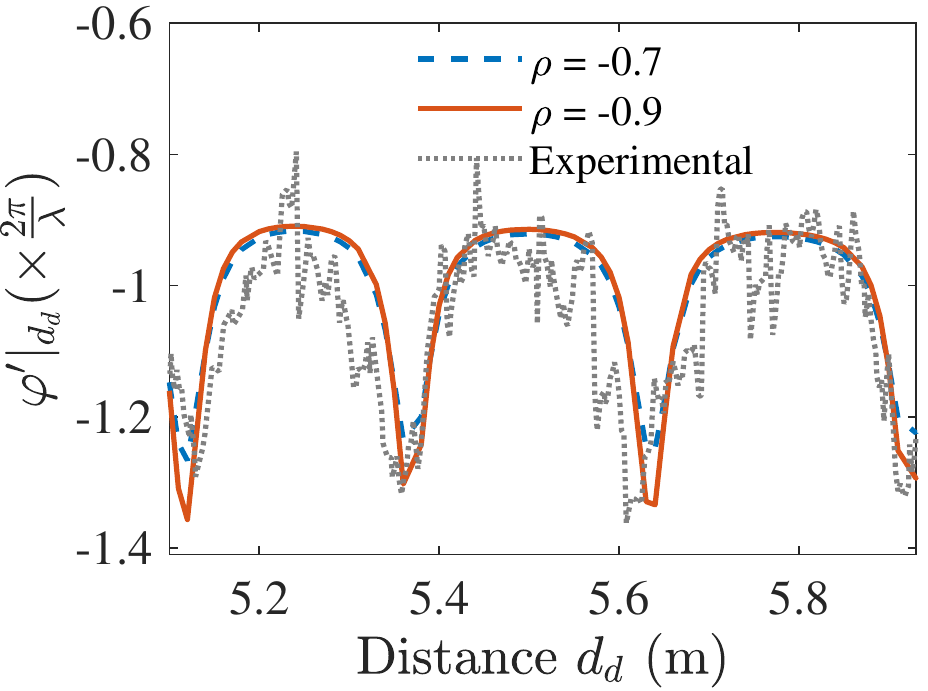}\label{daoshu_angle1}}
	\hfill
	\subfigure[Vertical flight]{
		\includegraphics[width=0.48\linewidth]{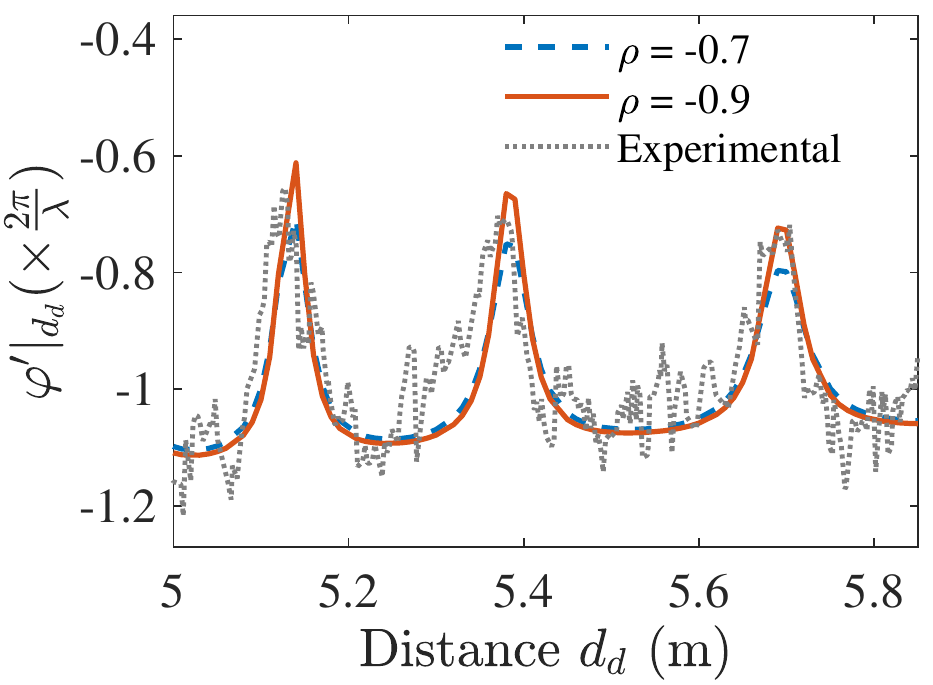}\label{daoshu_angle2}}
	\\
	\vspace{-0.2cm}\caption{Theoretical and experimental values of $\varphi'_{km}|_{d_d}$ when the UAV flies (a) horizontally or (b) vertically.}\vspace{-0.3cm}
	\vspace{-0.1cm}\label{daoshu_angle} 
\end{figure}
\begin{figure}[t]
	\centering
	\subfigure[Horizontal flight]{
		\includegraphics[width=0.44\linewidth]{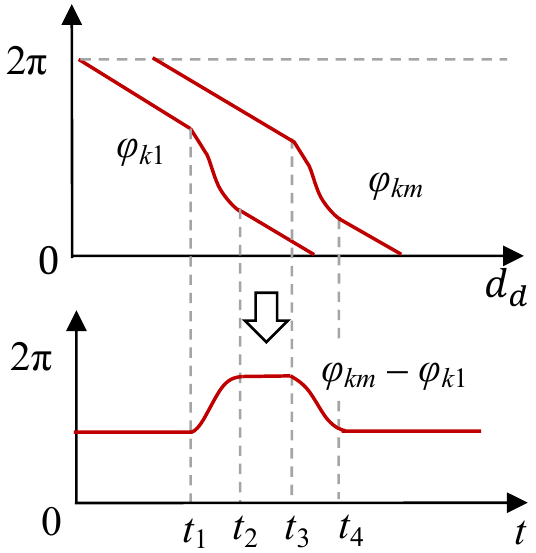}\label{angle_minus1}}
	\quad
	\subfigure[Vertical flight]{
		\includegraphics[width=0.44\linewidth]{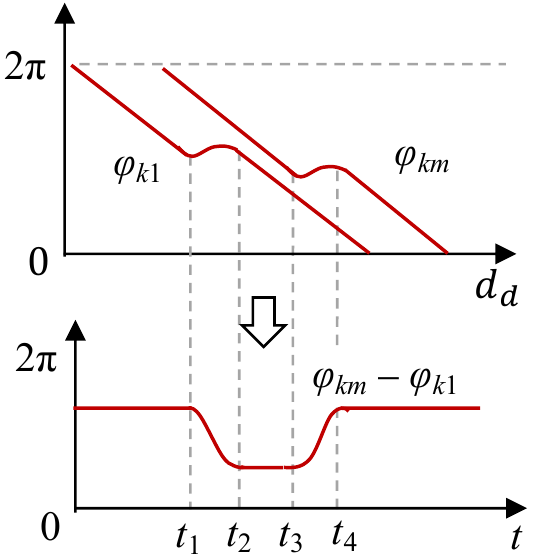}\label{angle_minus2}}
	\\
	\vspace{-1.5mm}\caption{Theoretical models of the changes in $\varphi_{k1}$,$\varphi_{km}$,$\Delta\varphi_{km}$ as the UAV flies (a) horizontally or (b) vertically in a short period.}
	\label{angle_minus} \vspace{-2mm}
\end{figure}

\begin{figure}[t]
	\centering
	\subfigure[Two-ray ground propagation]{
		\includegraphics[width=0.48\linewidth]{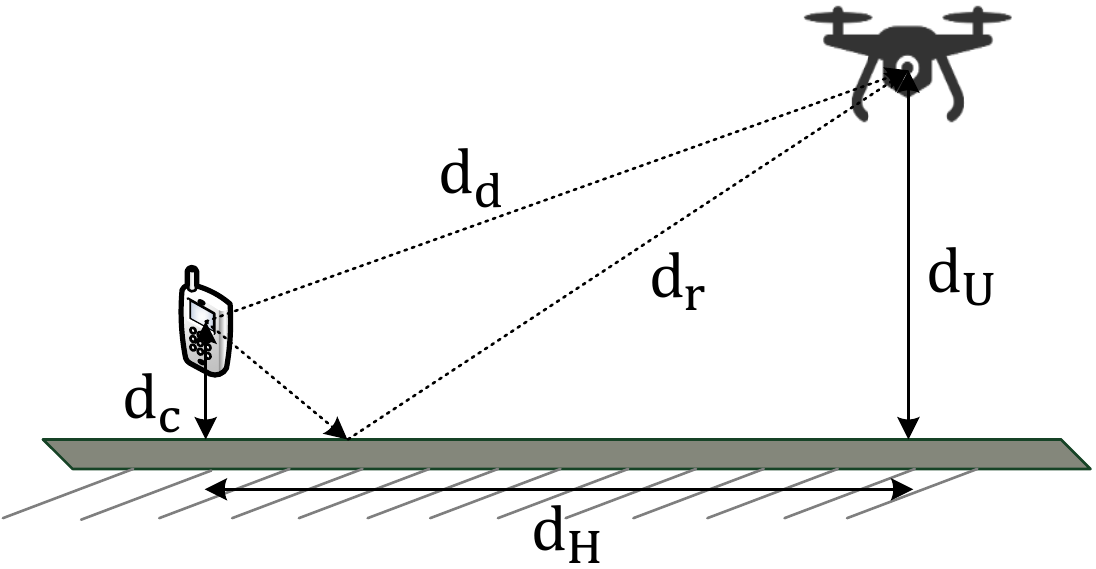}\label{two-ray}}
	\hfill
	\subfigure[UAV flight model]{
		\includegraphics[width=0.43\linewidth]{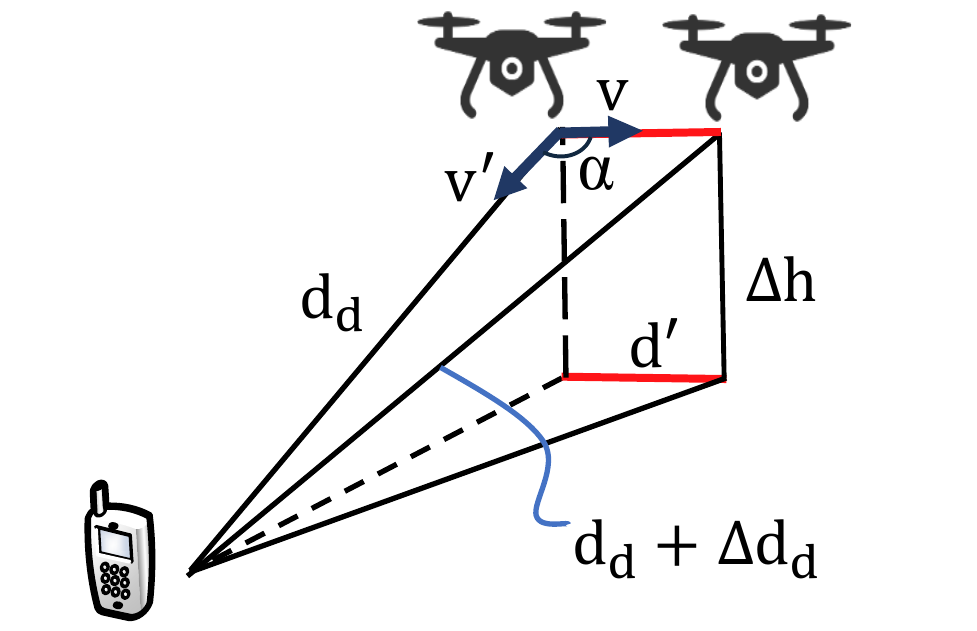}\label{sensor-assist}}
	\\
	\caption{Two-ray ground propagation model from a ground client to the UAV and the tree-dimensional UAV flight model.  }
	\vspace{-0.1cm}\label{uav-model} 
\end{figure}
\begin{figure}
	\centering
	\subfigure[]{
		\includegraphics[width=0.97\linewidth]{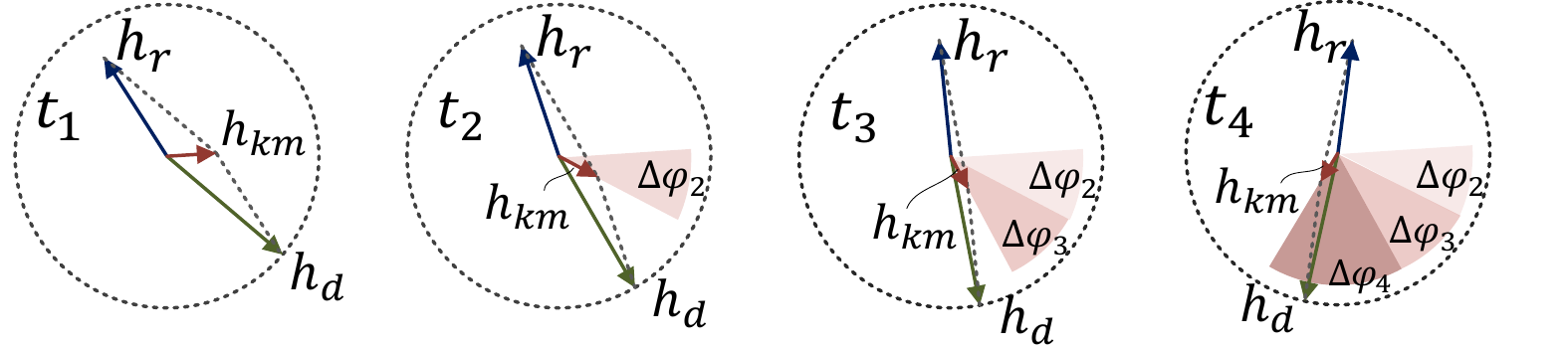}\label{angle_fading}}
	\hfill
	\subfigure[]{
		\includegraphics[width=0.97\linewidth]{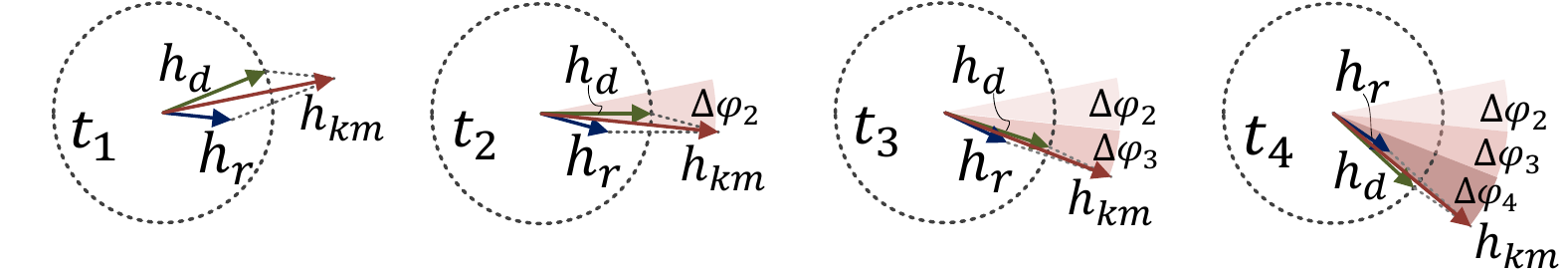}\label{angle_adding}}
	\\
	\vspace{-0.1cm}\caption{The dynamic changes in $\varphi_d$, $\varphi_r$, $\varphi_{km}$ as the UAV moves horizontally (a) during a fading period, $\Delta \varphi_2 \approx \varphi_3 \textless\ \varphi_4$; (b) during the period when $h_d$, $h_r$ interfere constructively, $\Delta \varphi_2 \approx \varphi_3 \approx \varphi_4$.}
	\label{angle_fading_adding} \vspace{-2.5mm}
\end{figure}

\textbf{Horizontal flight.} When the UAV flies in the horizontal direction, the hight of UAV $d_U$ is fixed and we assume the height of client $ d_c= 1$.  Then, $\gamma$ can be expressed by $\gamma=\frac{\sqrt{d_d^2+4d_U}}{d_d}$, and the derivation of $\gamma$ in terms of $d_d$ is $\gamma'|_{d_d}=\frac{-4d_U}{d_d^2\sqrt{d_d^2+4d_U}}$. 
Therefore, $a_0$ is computed as $a_0=\frac{d_d}{\sqrt{d_d^2+4d_U}}-1$. Due to the fact that $d_d\textgreater d_U-1$, $a_0$ is within the range of $a_0\in(\frac{-2}{d_U+1},0)$. Since  $\rho\in(-1,0)$, we can obtain the value of $\frac{a_0}{1-\frac{\gamma}{\rho}}\in(\frac{-1}{d_U+1},0)$ and  $\frac{a_0}{1+\frac{\gamma}{\rho}}\in(0,+ \infty)$. Thus, when $h_d$ and $h_r$ interfere constructively, $ \varphi_{km}'|_{d_d}$ is within a small range $(\frac{-2\pi}{\lambda}, \frac{-2\pi}{\lambda}(1-\frac{1}{d_U+1}))$, which remains relatively stable around $\frac{-2\pi}{\lambda}$, i.e., the phase changing speed of the direct-path link. In contrast, when $h_d$ and $h_r$ interfere destructively, $ \varphi_{km}'|_{d_d}$ belongs to a large range $(-\infty, \frac{-2\pi}{\lambda})$, which 
deviates a lot from $\frac{-2\pi}{\lambda}$. Fig.~\ref{daoshu_angle1} depicts both theoretical and experimental values of $ \varphi_{km}'|_{d_d}$ when the UAV flies in the horizontal direction. The experimental value fluctuates around the theoretical value, and the overall changing trend is similar.
Note that without fading, $ \varphi_{km}'|_{d_d}$ is relatively stable for large proportion of time. While fading occurs, $ \varphi_{km}'|_{d_d}$ suddenly decreases.
We plot in Fig.~\ref{angle_minus} the theoretical models of $\varphi_{km}$, $\varphi_{k1}$, $\Delta\varphi_{km}$. Fig.~\ref{angle_minus1} shows the case of horizontal flight. When no fading occurs, $\Delta\varphi_{km} =\varphi_{km}-\varphi_{k1}$ is basically unchanged because both $\varphi_{km}$ and $\varphi_{k1}$ have stable changing speed around $\frac{-2\pi}{\lambda}$. Nevertheless, when fading occurs for either $h_{k1}$ at time $t_1 $ or $h_{km}$ at time $t_3$ in turns, $\Delta\varphi_{km}$ experiences obvious increment or decrement under the multipath effect. We also depict in Fig.~\ref{uav-model} the plane and three-dimensional models of the UAV flight and radio propagation for better understanding.

Fig.~\ref{angle_fading} further shows one example of the dynamic changes in $\varphi_d$, $\varphi_r$ and $\varphi_{km}$ during the fading period. 
The green and blue arrows represent $h_d$, $h_r$ respectively, while the red arrow denotes the resulting $h_{km}$. We set the length of the arrow to amplitude and the direction to phase. Since $d_d$ and $d_r$ are unequal, the phase changing speeds are slightly different. From time $t_1$ to $t_4$, the phase difference between $\varphi_d$ and $\varphi_r$ gradually increases. 
When fading happens, 
their phase difference varies rapidly from $\textless180^\circ$ ($t_3$) to $\textgreater180^\circ$ ($t_4$). 
Thus, the direction of $h_{km}$ experiences a large-scale steering, from one side of the two arrows $h_d$, $h_r$ to the other side, much larger than the normal rotation speed. We can observe that $ \Delta \varphi_4\textgreater\Delta \varphi_3 \approx\Delta \varphi_2$ and $\varphi_{km}$ undergoes a more rapid changing process when fading occurs. 
When $h_d$ and $h_r$ interfere constructively (Fig.~\ref{angle_adding}), the  direction of $h_{km}$ also steers from one side to the other side at $t_3$-$t_4$. However, this steering is small, basically equal to the normal rotation speed. The changing speed of $\varphi_{km}$ doesn't vary too much.

\begin{figure}[t]
	\centering
	\includegraphics[width=0.5\textwidth]{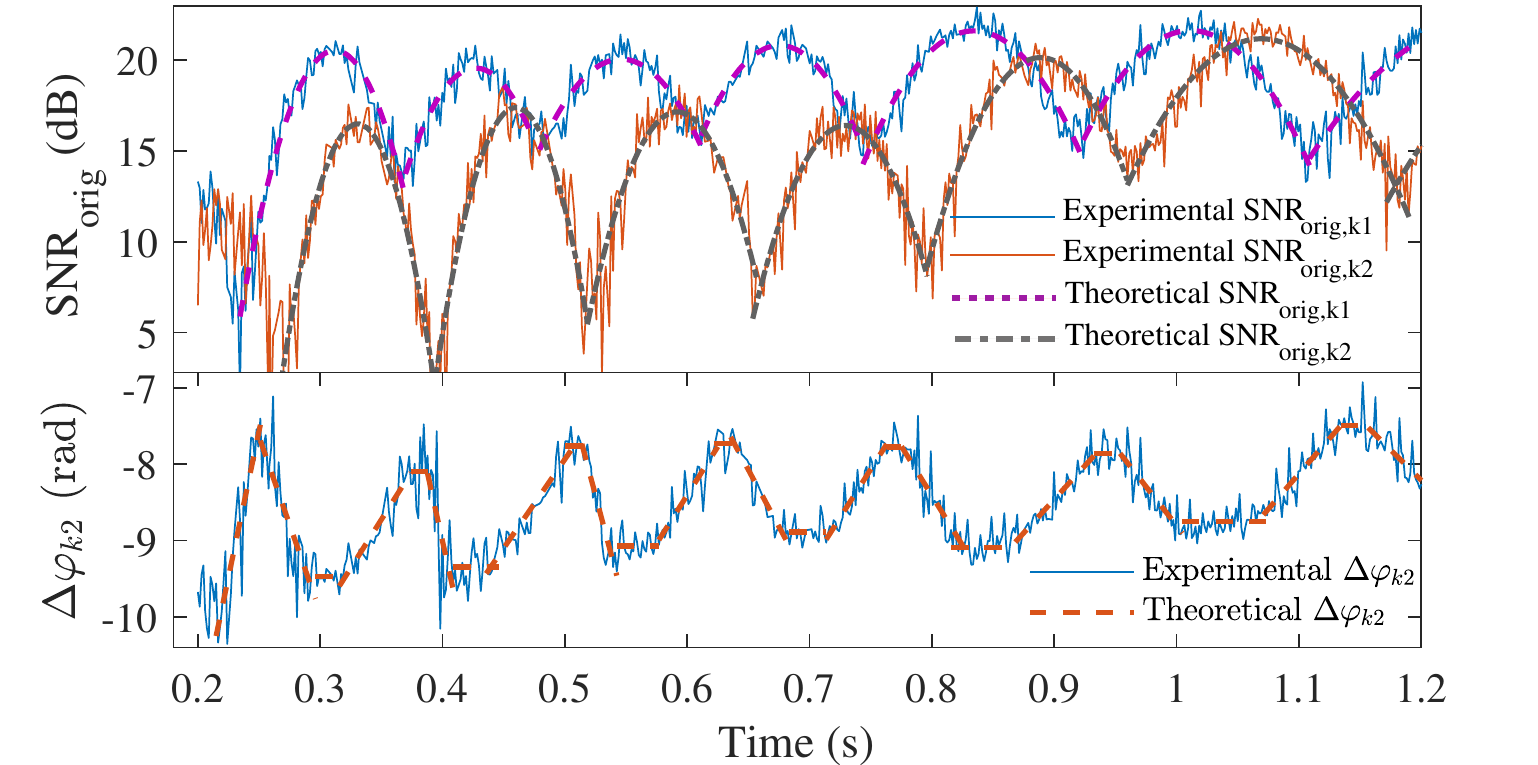}
	\vspace{-6mm}\caption{Experimental and theoretical values of $\text{SNR}_{orig,k1}$, $\text{SNR}_{orig,k2}$ and $\Delta \varphi_{k2}$ versus time when the UAV flies horizontally at variable velocities. }\vspace{-1mm}\label{zonghe}
\end{figure}

\textbf{Vertical flight.} When the UAV flies in the vertical direction, the horizontal distance $ d_H$ is fixed. Then,  $\gamma=\frac{\sqrt{d_d^2+4+4\sqrt{d_d^2-d_H^2}}}{d_d}$, and $a_0=\frac{2}{\gamma\sqrt{d_d^2-d_H^2}}+\frac{1}{\gamma}-1$. As the altitude of UAV gradually increases, we assume that $d_d$ satisfies the relationship $d_d \textgreater \sqrt{d_H^2+1}$. Thus, $a_0$ is within the range $a_0 \in (0, \frac{8}{\sqrt{1+9d_H^{-2}}*(3\sqrt{1+d_H^{-2}}+\sqrt{1+9d_H^{-2}})})$. When $d_H>1$, the range of $a_0$ can be simplified to $(0,0.34)$. Then, we can obtain the value of $\frac{a_0}{1-\frac{\gamma}{\rho}}\in(0,0.17)$ and $\frac{a_0}{1+\frac{\gamma}{\rho}}\in(- \infty,0)$. As a result, when $h_d$ and $h_r$ interfere constructively, $ \varphi_{km}'|_{d_d}$ is within a small range $(\frac{-2\pi}{\lambda}(1+0.17),\frac{-2\pi}{\lambda})$, close to the phase changing speed $\frac{-2\pi}{\lambda}$ of the direct-path link. In contrast, during a fading period, the value of $ \varphi_{km}'|_{d_d} $ belongs to a large range $(\frac{-2\pi}{\lambda}, +\infty)$. Fig.~\ref{daoshu_angle2} shows both theoretical and experimental values of $ \varphi_{km}'|_{d_d} $, which are consistent.
Fig.~\ref{angle_minus2} demonstrates the theoretical model in which $\Delta \varphi_{km}$ also decreases or increases significantly with fading occurrence for $h_{k1}$ or $h_{km}$, while remaining relatively stable for other conditions.

\textbf{Experimental validation.}
We evaluate the aerial channel changes on an open-air parking lot, where the channel measurements are affected by multiple reflectors, i.e., ground, trees and a car. 
Fig.~\ref{zonghe} shows both experimental and theoretical values of $\text{SNR}_{orig,k1}$, $\text{SNR}_{orig,k2}$ and $\Delta \varphi_{k2}$ when the UAV flies horizontally at variable velocities. 
The apparent reductions in $\text{SNR}_{orig}$ indicate the deep fades experienced by two channels $h_{k1}$, $h_{k2}$ in turn. 
When no fading occurs, $\Delta \varphi_{k2} $ remains basically stable, indicating that $\varphi'_{k1}|_d$ and $\varphi'_{k2}|_d$ are similar (and relatively stable around $\frac{-2\pi}{\lambda}$) during this period. 
When fading of $h_{k1}$ or $h_{k2}$ occurs in turn, $\Delta \varphi_{k2}$ increases or decreases as $\varphi'_{k1}|_d$ or $\varphi'_{k2}|_d$ deviates a lot from $\frac{-2\pi}{\lambda}$. 
These real-world measurements are consistent with the theoretical phase models in Fig.~\ref{daoshu_angle} and Fig.~\ref{angle_minus}, which gives us an opportunity to model the phase difference $\Delta\varphi_{km}$ changes as the UAV moves, based on the occurrence of deep fades.

\subsection{Sensor-Assisted Prediction Algorithm}\label{sec:design.c}

As theoretical models in Section~\ref{sec:design.b} show that $\Delta\varphi_{km}$ increases or decreases synchronously with the wireless fading, the fading forecast becomes paramount. However, due to the dynamic flight states, the fading positions and frequency vary significantly. 
Thus, we propose a channel prediction algorithm that exploits the flight-state-related sensor data to predict $\text{SNR}_{orig,k}$ and $\textbf{D}_k$ for each client $k$.

\begin{figure}
	\centering
	\includegraphics[width=0.5\textwidth]{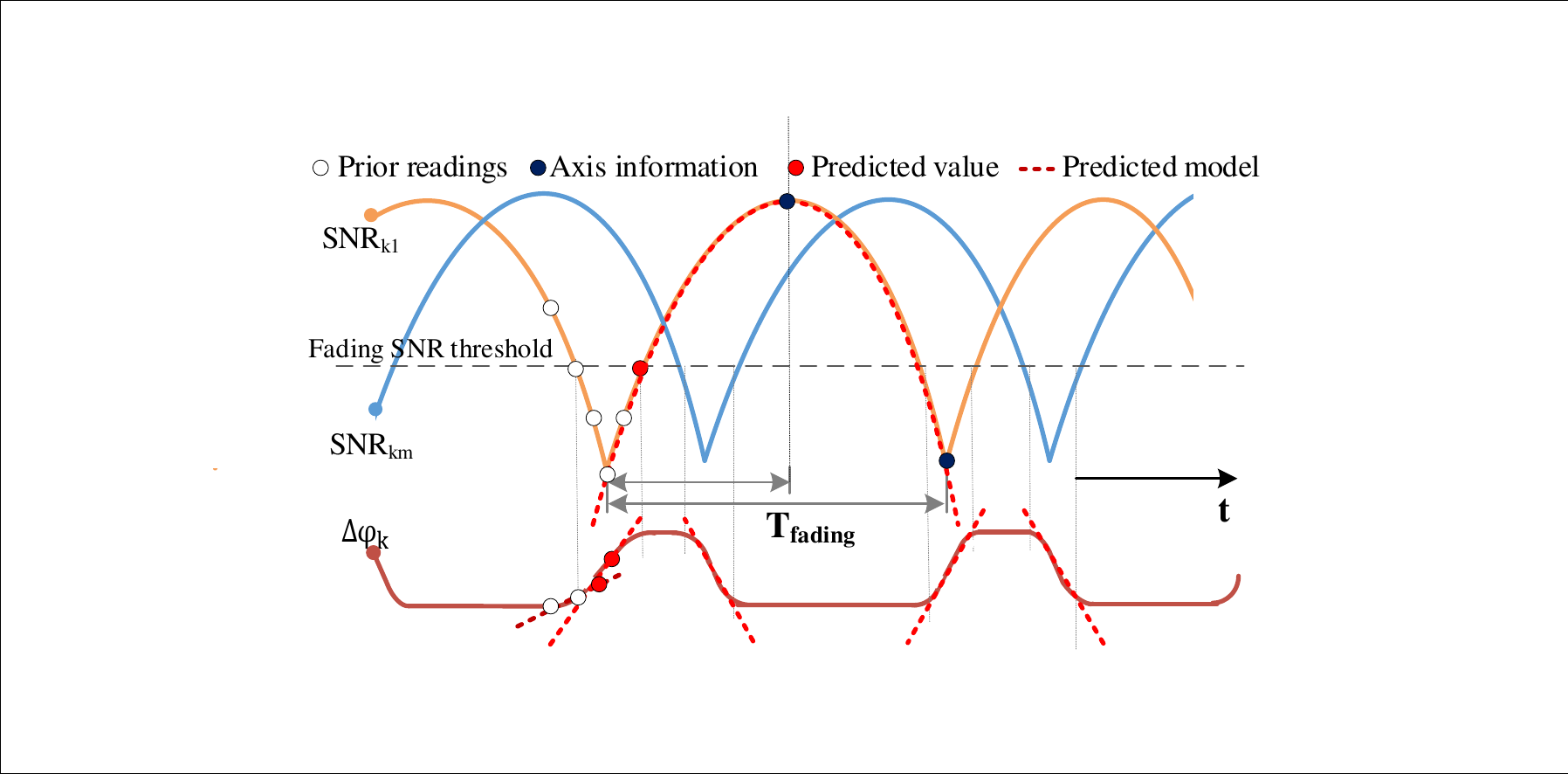}
	\vspace{-3.5mm}\caption{The prediction algorithm based on the past channel measurements and predicted fading frequency ($\frac{1}{T_{fading}}$).}\label{predict_analyze}\vspace{-3mm}
\end{figure}

\textbf{Prediction module.}
Fig.~\ref{predict_analyze} depicts the detailed prediction process. Inspired by \cite{chowdhery2018aerial}, we also adopt a second-order polynomial with time as the axis to fit and predict the fading patterns of each $\text{SNR}_{orig,km}$. As shown in Fig.~\ref{zonghe}, the experimental values of $\text{SNR}_{orig,km}$ fit the second-order polynomial curve. However, different from the constant velocity of 1~m/s in \cite{chowdhery2018aerial}, the flight states are highly dynamic in our experiment. This greatly affects the channel variance, especially the fading frequency $f_{fading}$ ($T_{fading} = \frac{1}{f_{fading}}$, shown in Fig.~\ref{predict_analyze}).
To tackle this problem, we predict $f_{fading}$ as a function of UAV's relative velocity $\textbf{v}$ and position $\textbf{p}$ to the client, i,e., $f_{fading}=g(\textbf{v}, \textbf{p})$. This function provides the information of axis of symmetry for the second-order polynomial to better adapt to the dynamic flight states. In addition to the axis information, SensRate takes the past $\text{SNR}_{orig,km}$ measurements to form a set of points (blue and white points in Fig.~\ref{predict_analyze}) that build the second-order polynomial regression. This polynomial will be updated as new $\text{SNR}_{orig,km}$ becomes available and thoroughly initialized after detecting the lowest $\text{SNR}_{orig,km}$. Based on this second-order polynomial function, we can predict $\text{SNR}_{orig,km}$ change in the near future.

In order to predict $\textbf{D}_k$ changes, we need to accurately detect the beginning and the end of a fade because $\Delta\varphi_{km}$ changes significantly during the fading period. Typically, the SNR during a fade is lower than the value before and after the fading. So we set an SNR threshold to indicate the start and the end of a fade. This threshold is empirically set as the medium value of past several SNR fluctuation ranges. 
Once detecting that the newly predicted $\text{SNR}_{orig,km}$ or $\text{SNR}_{orig,k1}$ is less than their respective thresholds, we adopt a linear function that relies on the past $\Delta\varphi_{km}$ measurements to fit the changes in $\Delta\varphi_{km}$ and predict the future value, as shown in Fig.~\ref{predict_analyze}. 
From the start to the end of a fade, this linear function is updated once new $\Delta\varphi_{km}$ measurement is available. After the fading ends, the phase difference becomes relatively stable and we can directly utilize the last measured value. 
It is noteworthy that when both channels $h_{km}$, $h_{km}$ are during the fading period, the phase difference remains relatively stable, since the two channels have similar phase changing trends.

After predicting the changes in $\text{SNR}_{orig,km}$, $\text{SNR}_{orig,k1}$ and $\Delta\varphi_{km}$, we can calculate each $h_{km}/h_{k1}$ that forms $\textbf{D}_k$, where $|h_{km}|$ is converted from $\text{SNR}_{orig,km}$. Next, each client can predict its channel direction $\textbf{D}_k$ and estimate the angle $\theta_k$ between itself and the already ongoing clients. Additionally, the $\text{SNR}_{orig,k}$ can be calculated by $\text{SNR}_{orig,km}, m=1,2,..M$. Then, we can obtain the resulting $\text{SNR}_{proj,k}$ according to Eq.~\eqref{sinr}-\eqref{sinr2} for each client to select the transmission rate.

\textbf{Predicting fading frequency using sensor data.}
In this part, we describe the detailed prediction process of fading frequency $f_{fading}=g(\textbf{v},\textbf{p})$ that relies on the relative velocity $\textbf{v}$ and position $\textbf{p}$ of the UAV to each ground client.
The UAV periodically broadcasts its 3D velocity and 3D position. If the ground client is equipped with IMU and GPS, it can directly calculate $\textbf{v}$ and $\textbf{p}$. However, for the client without IMU or GPS, we can further employ the technique in \cite{10.1145/3411833} by leveraging the time of flight (ToF) and Doppler shift to estimate the velocity $v'$ in the direct-path direction and the UAV-to-client distance $d_d$. In this case, the relative position $\textbf{p}$ of the UAV can be obtained when ignoring the change in height $d_c$ of the client. 

\begin{figure}
	\centering
	\subfigure[$T_{fading}$ vs. velocity.]{
		\includegraphics[width=0.48\linewidth]{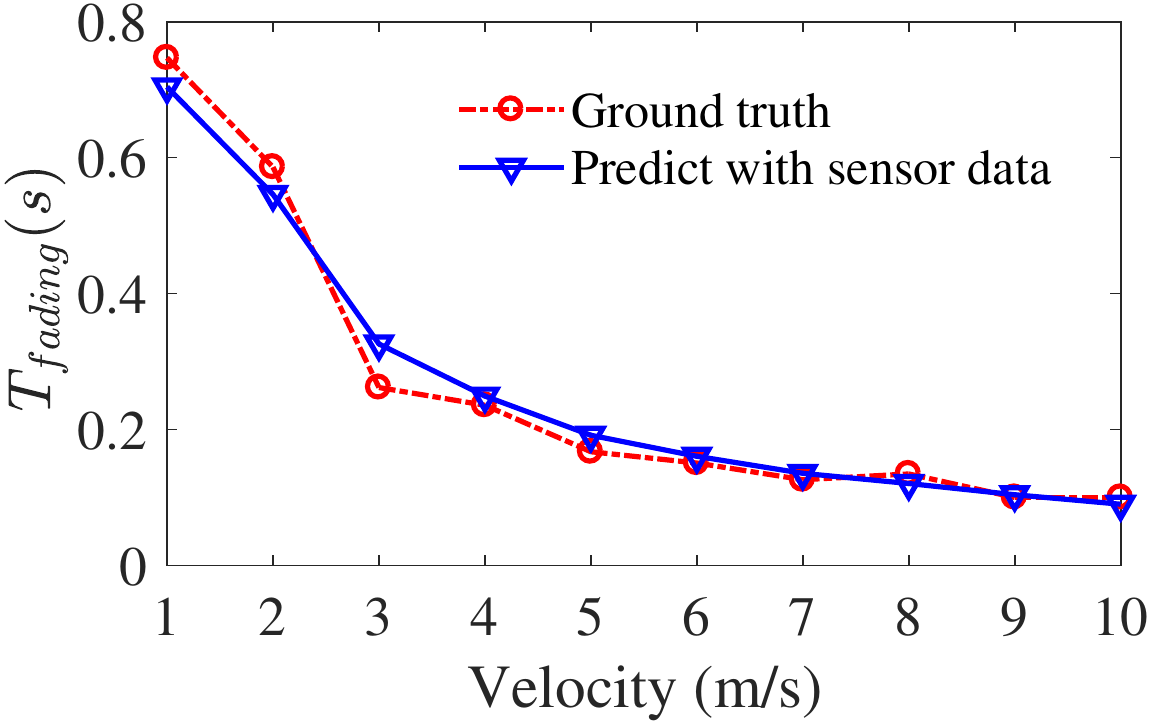}\label{fadingf_v}}
	\hfill
	\subfigure[$T_{fading}$ vs. distance.]{
		\includegraphics[width=0.48\linewidth]{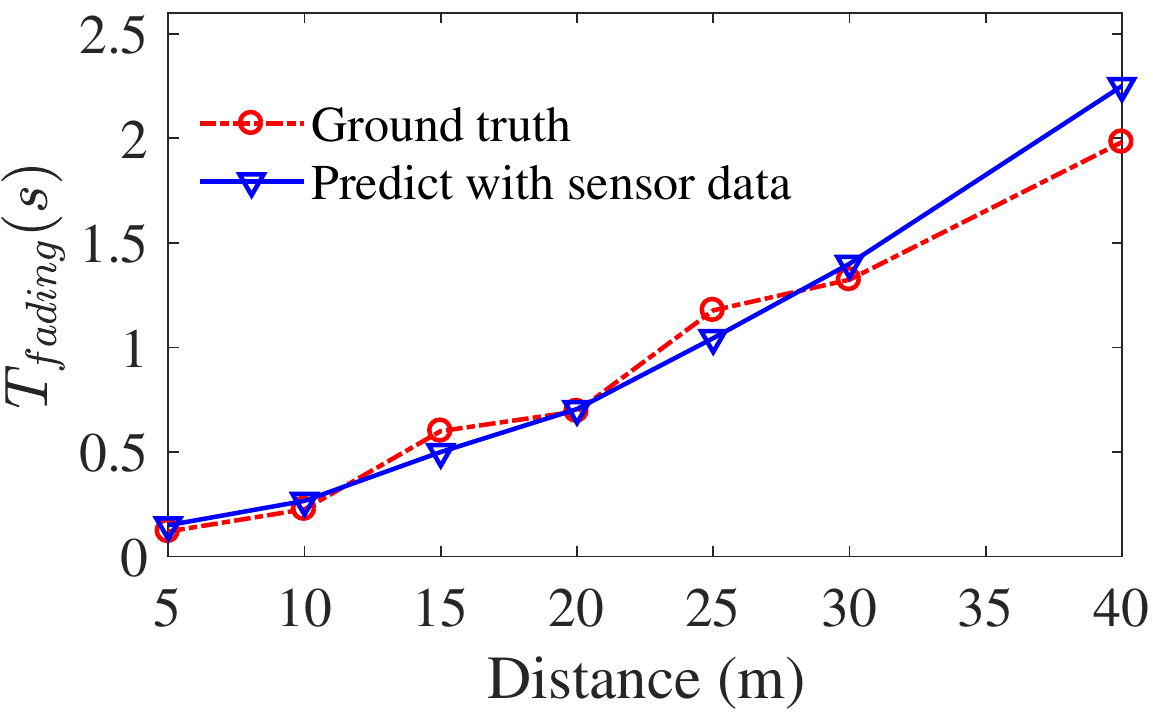}\label{fadingf_d}}
	\\\vspace{-1mm}
	\caption{The average $T_{fading}$ with various UAV velocities or UAV-to-client distances, including the predicted value and the ground truth. } \vspace{-2.5mm}
	\label{fading_f} 
\end{figure}
Recall that we can estimate the direct-path distance $d_{d,\beta}$ corresponding to each fading position as $d_{d,\beta} =\frac{\beta\lambda}{(\gamma-1)},\beta\in \mathbb{Z} $. For each $\textbf{p}$ along UAV trajectories, there are two representations of $\gamma$ in the horizontal or vertical direction, which are
\begin{equation}\label{gamma}
	\gamma=
	\begin{cases}
		\frac{\sqrt{d_d^2+4d_Ud_c}}{d_d}, & d_U, d_c \  \text{as}  \ \text{ parameters},\\
		\frac{\sqrt{d_d^2+4d_c^2+4d_c\sqrt{d_d^2-d_H^2}}}{d_d},& d_H, d_c \  \text{as}  \ \text{ parameters}.
	\end{cases}
\end{equation}
When estimating $d_{d,\beta}$ in the horizontal direction, $d_U, d_c$ are as known parameters with only $d_d$ as a variable. The estimation of $d_{d,\beta}$ in the vertical direction is similar. Thus, we can solve two different series of $\{d_{d,\beta}\}s, \beta\in \mathbb{Z}$ for each $\textbf{p}$.
If the current real-time \textbf{v} is exactly close to the vertical or horizontal direction,
we calculate the real-time $d_d$ and search its interval $d_{d,\beta-1}\leq d_d\textless d_{d,\beta}$ along this direction. The change in direct-path distance $\Delta d_d$ between adjacent fading positions can be computed by $\Delta d_d= d_{d,\beta}- d_{d,\beta-1}$.

As shown in Fig.~\ref{sensor-assist}, we convert the absolute value of relative velocity $v = |\textbf{v}|$ into the velocity $v'$ in the direct-path direction by
\begin{equation}\label{v'}
	v' = v*|\cos\alpha|,
\end{equation} 
where $\alpha$ is the angle between the relative velocity of UAV and the direct-path link, which can be directly calculated through $\textbf{v}$ and $\textbf{p}$.
The fading interval $T_{fading}$ can be predicted by 
\begin{equation}\label{T_fading}
	\Delta d_d =  \begin{matrix} \int^{T_{fading}}  v*|\cos\alpha| \,  dt\end{matrix}.
\end{equation}
However, if the current real-time \textbf{v} is oblique, the
$T_{fading}$s in vertical direction and horizontal direction should be separately estimated, and the smaller value is selected. It is noteworthy that $T_{fading}$ should be continuously updated especially when there exits oblique flight in UAV trajectory, because $d_U$ or $d_H$ is not constant in this case.

We conduct real-world experiments on an open-air parking lot to compare the predicted $T_{fading}$ with the ground truth. Fig.~\ref{fading_f} shows the average $T_{fading}$ in horizontal flights with various UAV velocities or UAV-to-client distances. 
When testing $T_{fading}$ vs. UAV velocity, the UAV-to-client distance is limited within 10-20~m. When testing $T_{fading}$ vs. distance, the velocity is limited within 2-3~m/s.
Note that $T_{fading}$ decreases with the velocity and increases with the distance, which is in line with Eq.~\eqref{gamma}-\eqref{T_fading}.
We can observe that the predicted values are similar to the ground truth, verifying that SensRate does perform well in adapting to different flight states.

\begin{figure*}[t]
	\centering
	\subfigure[]{
		\includegraphics[width=0.255\linewidth]{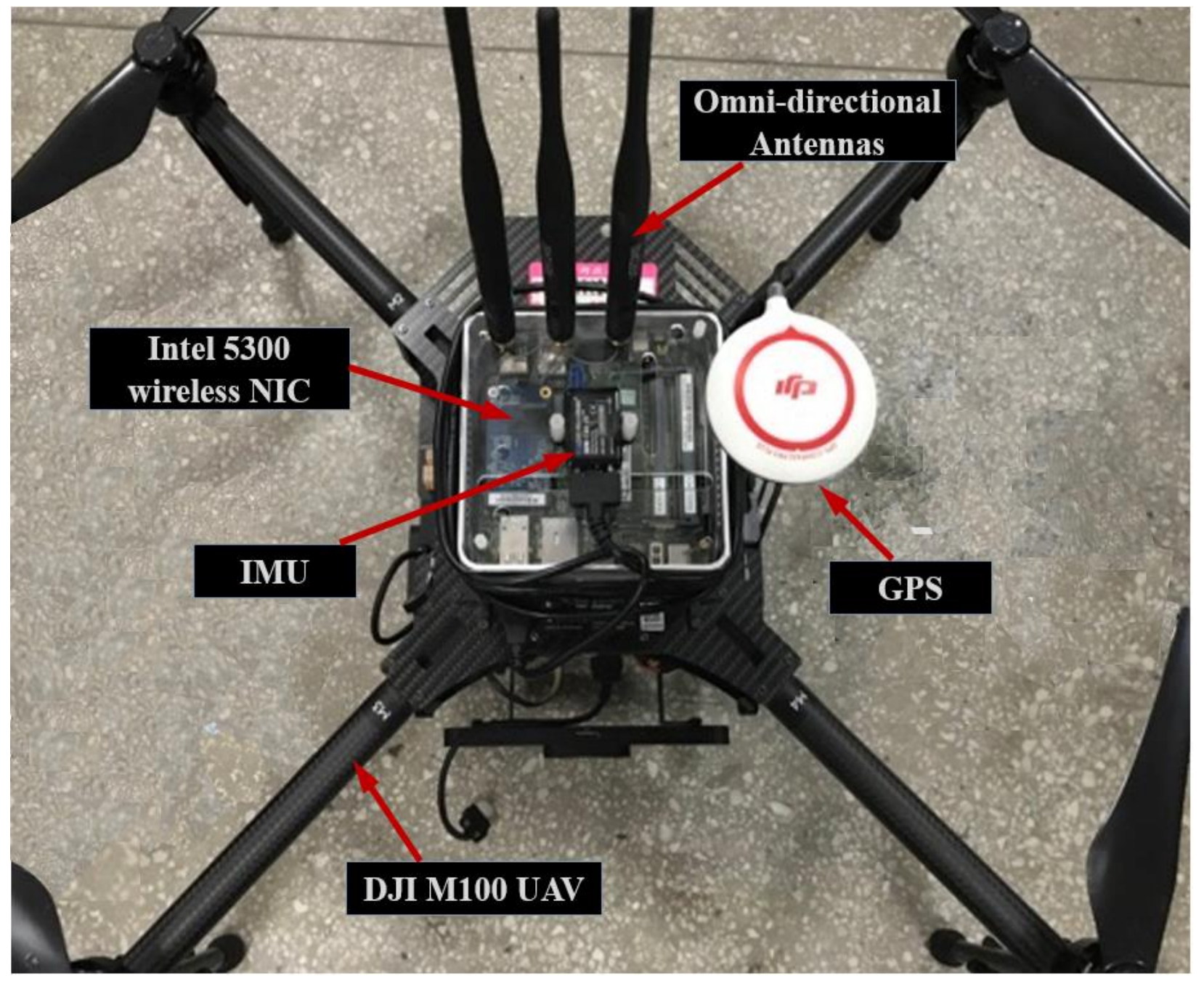}\label{uav1}}
	\hfill
	\subfigure[]{
		\includegraphics[width=0.1715\linewidth]{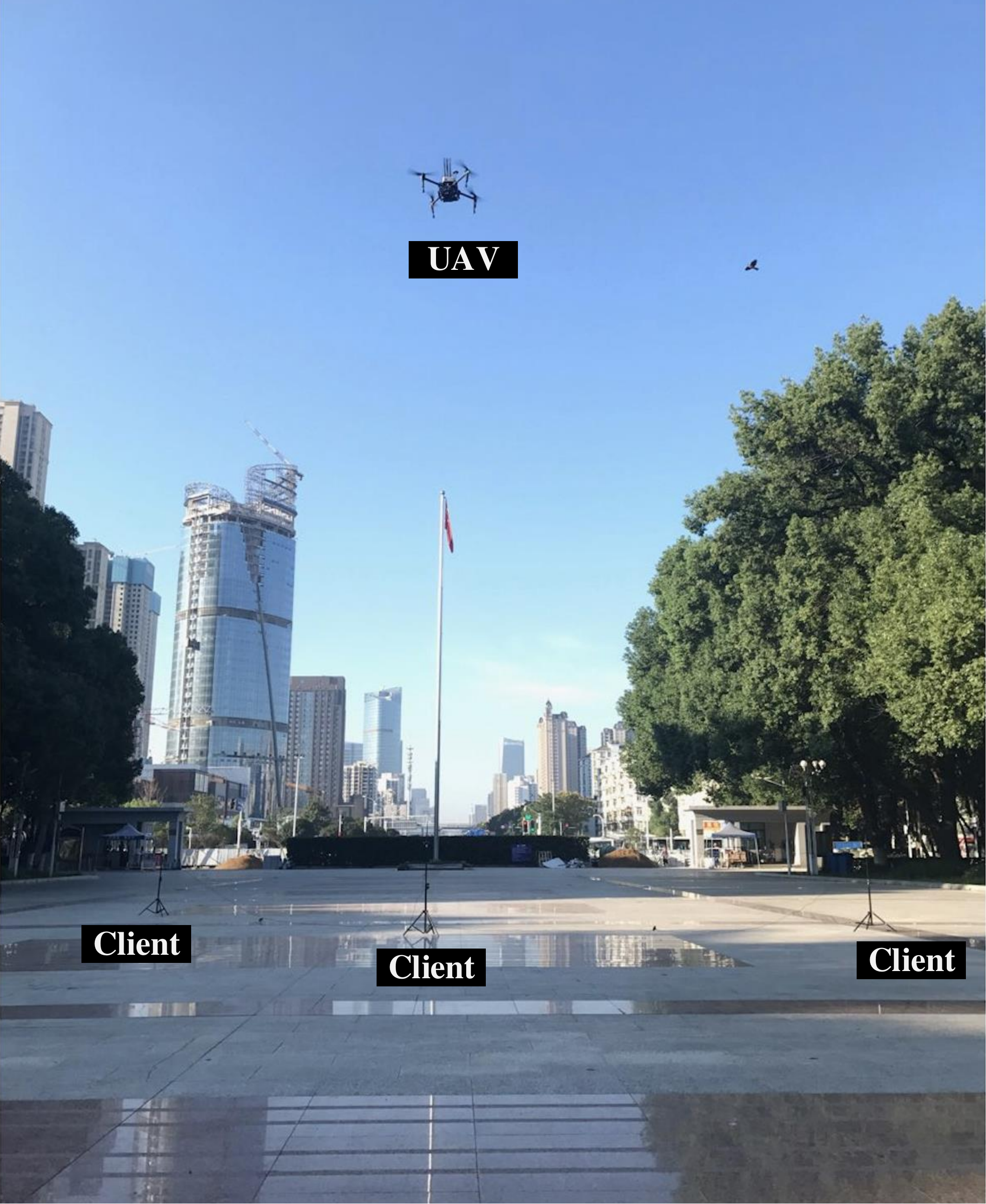}\label{square}}
	\hfill
	\subfigure[]{
		\includegraphics[width=0.171\linewidth]{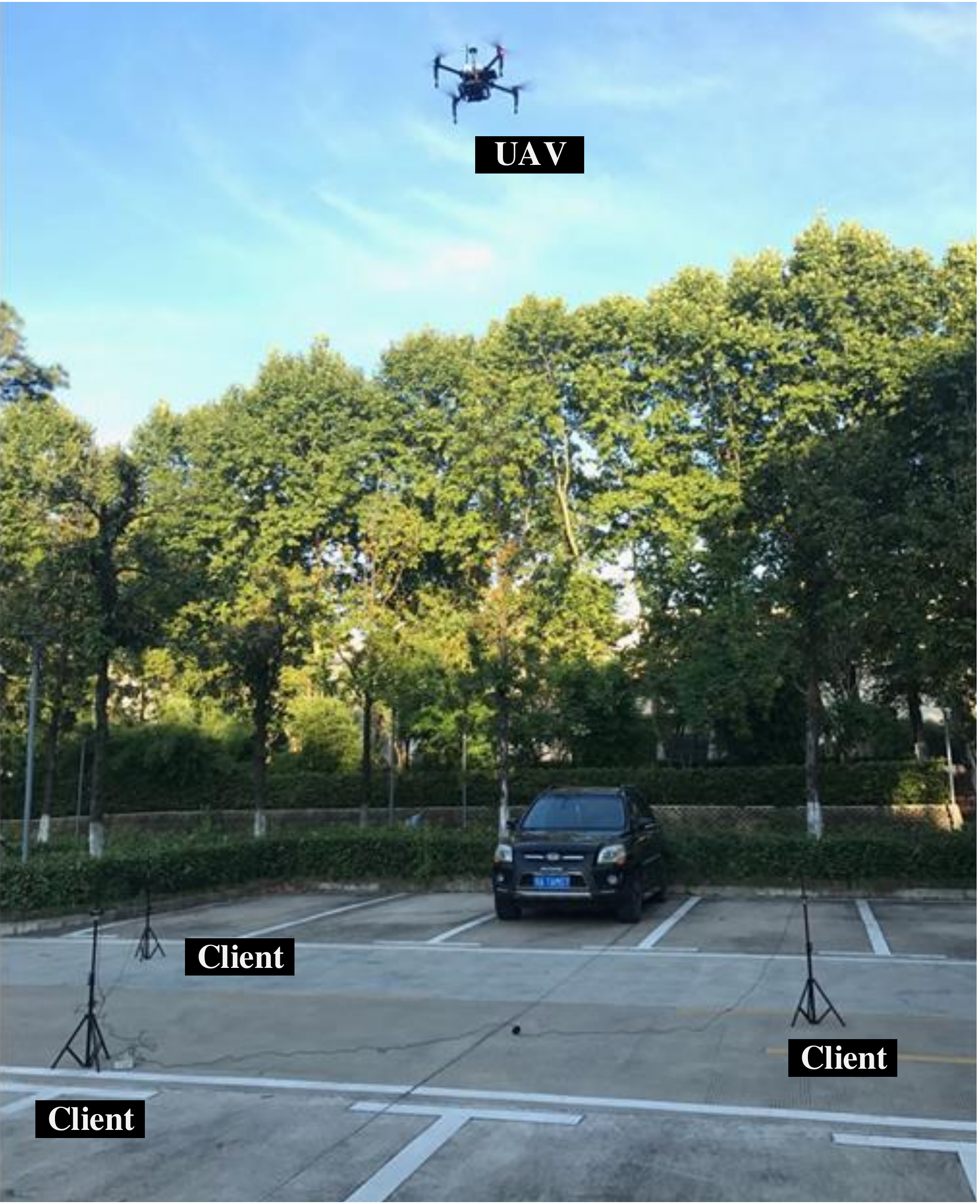}\label{parking}}
	\hfill
	\subfigure[]{
		\includegraphics[width=0.273\linewidth]{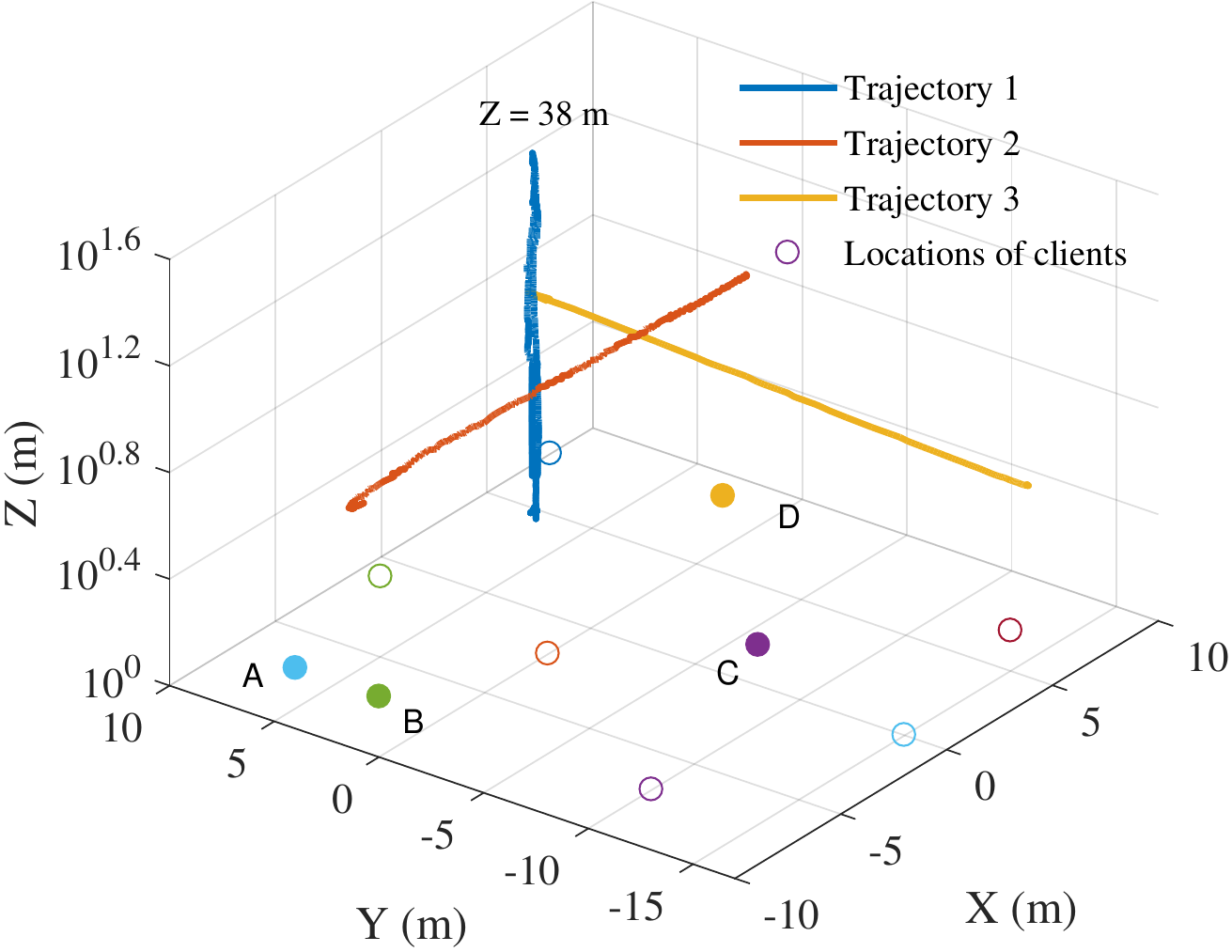}\label{trajectory_3d}}
	\\
	\vspace{-0.1cm}\caption{Experimental setup. (a) Implementation. (b) An experimental site: an empty square. (c) Another experimental site: an open-air parking lot with a car on it and trees surround it. (d) The map of UAV trajectories and clients' locations shared by both experimental sites.}\label{uav}\vspace{-0.2cm}
\end{figure*} 

\subsection{Rate Selections and ZF-SIC Decoding}

According to Section \ref{sec:design.a}, clients join transmissions one after another (say from client 1 to $K$). 
Client 1 can determine the rates only based on its predicted $\text{SNR}_{orig}$. 
Later clients estimate the inter-user interference and $\text{SNR}_{proj}$ for rate selections. SensRate uses zero-forcing with successive interference cancellation (ZF-SIC)~\cite{turborate,tse2005fundamentals} to decode in descending order from client $K$ to 1. Specifically, once the $k_{th}$ stream is successfully decoded by projecting the received signal along the direction orthogonal to the subspace consisting of $\textbf{D}_1$ to $\textbf{D}_{k-1}$, we subtract it from the received signals. 

It is noteworthy that there exists error propagation and imperfect interference cancellation in ZF-SIC decoding. 
By applying the technique in \cite{lin2011random,turborate} to cope with the imperfect cancellation, each client compares its predicted $\text{SNR}_{orig}$ with a threshold of 25-27~dB before transmission. If the $\text{SNR}_{orig}$ exceeds this threshold, the client (except client 1) will reduce the transmission power to avoid the residual noise after cancellation. 
Besides, due to the error propagation, the BER of client $k$ will be affected by whether the previous streams ($(k+1)_{th}$-$K_{th}$) are decoded correctly. 
According to~\cite{tse2005fundamentals}, we can estimate the error probability of decoding the stream from client $k$ (for simplicity, we treat it as $\text{BER}$ $p_k$) by
\begin{equation}\label{error probability}
	p_k =p_{k+1}+(1-p_{k+1})*p_e^{k}=\sum_{i=k}^{K}\left(p_e^{i}\prod_{j = k+1}^{K}(1-p_e^{j})\right),
\end{equation}
where 
$p_e^{i}$ is the BER of decoding the $i_{th}$ stream incorrectly when all the previous streams are decoded correctly. 
Then we calculate the effective $\text{SNR}_{eff,i}$ 
and estimate the coded bit error rate $p_{e,R_i}^{i}$ of 
each rate selection $R_i$, based on the predicted subcarrier $\text{SNR}_{proj,i}$. Then, by combining $p_{e,R_i}^{i}$ with Eq.~\eqref{error probability}, the throughput $\eta_{k}$ of client $k$ can be estimated as
\begin{equation}\label{eta}
	\eta_{k} = R_k(1-p_{k})^L= (1-p_{k+1})^L*R_k(1-p_{e,R_k}^{k})^L,
\end{equation}
where $L$ is the number of bits in a packet.

The aim of SensRate is to schedule proper $R_k$ for all concurrent clients to maximize the overall throughput (formulated as Eq.~\eqref{overall_throughput}).
A key point is that the AP decodes the streams in order of client $K$ to 1. We can observe from Eq.~\eqref{eta} that the $p_{k+1}$ of the previously decoded streams will not affect the rate selection $R_k$ of client $k$ to maximize its throughput $\eta_{k}$. But the rate selection $R_k$ of each client $k$ and its resulted $p_{e,R_k}^{k}$ will affect the decoding of remaining streams ($1_{th}$ to $(k-1)_{th}$). Thus, the multi-client rate selections for Eq.~\eqref{overall_throughput} can be converted to the rate selection of each client $k$ as
\begin{align}\label{maximize}
	&\max_{R_1,\cdots,R_K} \sum^K \eta_{k}\rightarrow \max_{R_k}\sum_{i=1}^k \eta_i, \quad k = 1,2, \cdots,K \notag\\
	&= \max_{R_k} \left[ (1-p_{k+1})^L*\sum_{i=1}^k \left(R_i*\prod_{j = i}^k(1-p_{e,R_j}^{j})^L\right)\right].
\end{align}
Limited by the inherent problem that client $k$ doesn't know the conditions (SNR, $R$, $p_e$, etc.) of other clients, it assumes the conditions of later clients are as same as itself to obtain the $R_k$ of Eq.~\eqref{maximize}, which is simplified to  
\begin{equation}\label{maximize_simplify}
	\max_{R_k} \left[ (1-p_{k+1})^L*R_k*\sum_{i=1}^k \left(\prod_{j = i}^k(1-p_{e,R_k}^{k})^L\right)\right].
\end{equation}
This module enables each client $k$ to select the proper rate $R_k$ over all available physical-layer rates to maximize the overall throughput in a fully distributed manner.

\subsection{Additional Issues}
\textbf{Decoding scheme}.
We choose ZF to decode the concurrent streams due to its relatively low complexity and good performance in the high SNR regime. Since each client can estimate the $\text{SNR}_{proj}$ before joining the contention, SensRate forces clients whose estimated $\text{SNR}_{proj}$ are below 4~dB (minimum SNR for rate selection~\cite{turborate}) to give up joining the contention, which avoids the performance degradation of ZF in the low SNR regime and maximizes the overall throughput.

\textbf{None-line-of-sight (NLOS).}
As stated in \cite{6863654}, compared to LOS propagation, an extra path loss of 20-30~dB needs to be further subtracted in NLOS propagation, which are measured in Suburban, Urban, Dense Urban, and Highrise Urban. Thus, compared to clients under LOS conditions, the clients under NLOS conditions have much lower probabilities of joining transmission at the current moment. Instead, when the UAV patrols around, they join the transmission when the LOS propagation path becomes available.

\textbf{Overhead.}
We finally check the extra overhead introduced by SensRate. The overhead mainly comes from (1) the UAV broadcast packets, and (2) the exchange of channel directions among clients. 
As the experimental results shown in Section~\ref{channel_prediction}, the UAV broadcast rate of 50-100~Hz is required for SensRate prediction. The UAV broadcast packet includes 3D positions ($3\times4$~bytes) and 3D velocities ($3\times4$~bytes).
The transmission time is $(24*8/6e6 +100)\times50(100) = 6600~\mu s(13200~\mu s)$, which only requires 0.66\%-1.32\% of the available airtime at the lowest rate of 6~Mbps. The 100~$\mu s$ comes from the PLCP header that contains the training preamble. 
Like TurboRate~\cite{turborate}, we also transform the
channel directions across all subcarriers to the time domain. The clients only broadcast the
first few significant taps, e.g., five taps, of the time-domain channels, resulting in a 4\% average throughput loss~\cite{turborate} as an inevitable sacrifice. In addition, the computational cost at the UAV side mainly comes from the ZF-SIC decoding, with the complexity of $\mathcal{O} (KM^3)$. The cost at the client side includes (1) the phase difference prediction ($\mathcal{O} (KM)$), (2) the $\text{SNR}_{orig}$ prediction ($\mathcal{O} (KM)$), (3) the $\theta$ calculation ($\mathcal{O} (KM^3)$), (4) the $\text{SNR}_{proj}$ calculation ($\mathcal{O} (K)$),  and (5) the rate selection ($\mathcal{O} (K)$), with total complexity of $\mathcal{O} (KM^3)$. Therein, we simplify the cost of linear/curve fitting to $\mathcal{O}(1)$, which originally relates to the number of samples.

\section{Performance Evaluation}\label{sec:evaluation}
\subsection{Experimental Setup}\label{sec:evaluation.a}
\textbf{Implementation.}
We implement SensRate on a commercial UAV platform, DJI Matrice 100, which serves as a mobile hotspot to communicate with several ground clients. As shown in Fig.~\ref{uav1}, the UAV is equipped with sensors including inertial measurement unit (IMU), GPS receiver, barometer, magnetometer and ultrasound to obtain its real-time 3D positions and 3D velocities. Besides, an Intel next unit of computing (NUC) with an Intel 5300n wireless chipset is equipped on the UAV, which is connected with three omni-directional antennas to act as a multiple-antenna AP. On the ground, we deploy a controller NUC equipped by an Intel 5300n wireless chipset to connect with three omni-directional antennas. By placing the antennas separately at different positions, we regard them as several independent clients. 
The wireless cards work in the monitor mode, operating on a 20 MHz channel at 5~GHz frequency band. We leverage the Intel 5300 CSI tool~\cite{csitool} to record air-to-ground CSI at 1000~Hz as well as corresponding timestamps. 
The collected CSI traces are then downsampled to different CSI reading rates to evaluate the performance of SensRate.
Although this CSI is measured in a 3$\times$3 MIMO mode, it can evaluate advanced techniques like multi-user MIMO~\cite{csitool} in our experiment, where each antenna on the ground acts as a client. 

The NUC on the UAV logs the timestamps, 3D positions and 3D velocities at 50~Hz. 
To minimize the time offset between NUCs, we connect them to the campus network to synchronize their system time before each UAV flight. As the real-time clock (RTC) drift of Intel NUC is 24~ppm~\cite{intel-nuc}, the time offset between NUCs during each flight is limited and do not affect our experiments. Thus, we use the timestamps to align the sensor data and CSI. It is noteworthy that this alignment is not required in our system design, as clients can measure CSI and obtain corresponding UAV sensor data from UAV broadcast packets without additional alignment.
The experimental sites include an empty square and an open-air parking lot with a car on it and trees surround it, as shown in Fig.~\ref{square} and Fig.~\ref{parking}. 
The UAV is controlled to fly in different trajectories with varying flight states according to  experimental requirements.
We use the collected traces to evaluate SensRate's performance on top of an 802.11-compatible MU-MIMO OFDM library~\cite{signpost}. To realize SensRate, we add the function of sensor-assisted channel prediction and rate adaptation to this MU-MIMO communication system.

\textbf{Baseline algorithms.} 
For performance comparison, we have also implemented three baseline RA algorithms: 
\begin{itemize}
\item \textit{TurboRate}~\cite{turborate} is the state-of-the-art RA algorithm for UL MU-MIMO, which takes into account the inter-user interference as a bias for transmission rate selections. Clients in TurboRate exchange the past channel direction measurements to calculate the inter-user interference. 

\item \textit{OPT} is an omniscient algorithm that provides upper bounds for all the algorithms. It knows in advance the CSI, the inter-user interference and the $\text{SNR}_{proj}$ for all clients, and ensures the highest transmission rate for the next frame that can be successfully sent.

\item \textit{ESNR}~\cite{Predictable} refers to the single-user RA algorithm, which selects the optimal 802.11n MIMO rates based on effective SNR of past channel information. Only one client is allowed to transmit to a multi-antenna AP at a time  through traditional 802.11 content mechanism. 

\end{itemize}

\begin{figure}[t]
	\centering
	\subfigure[Throughput for each client in the 2-antenna AP scenario.]{
		\includegraphics[width=0.48\linewidth]{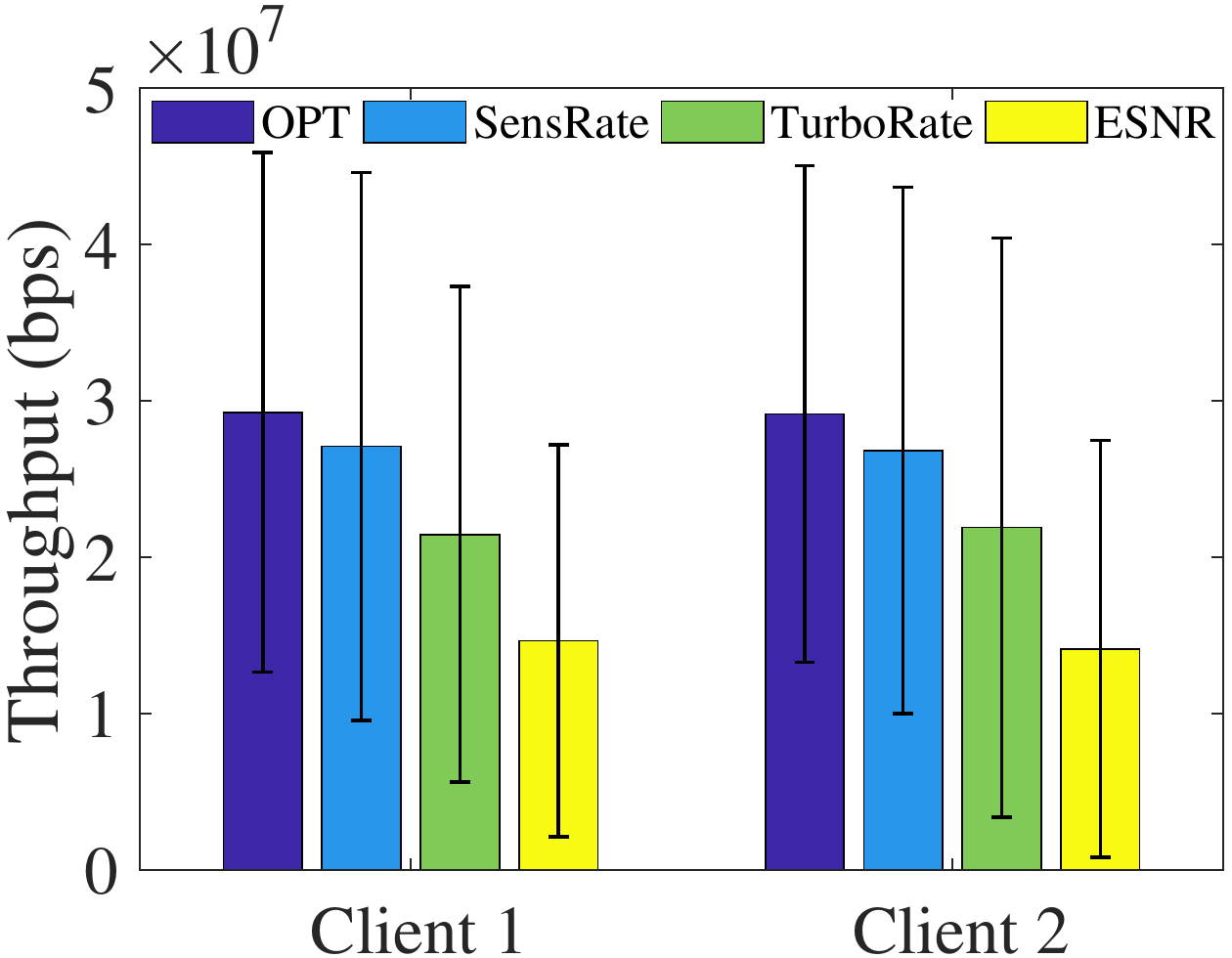}\label{2an-th}}
	\hfill
	\subfigure[Throughput for each client in the 3-antenna AP scenario.]{
		\includegraphics[width=0.48\linewidth]{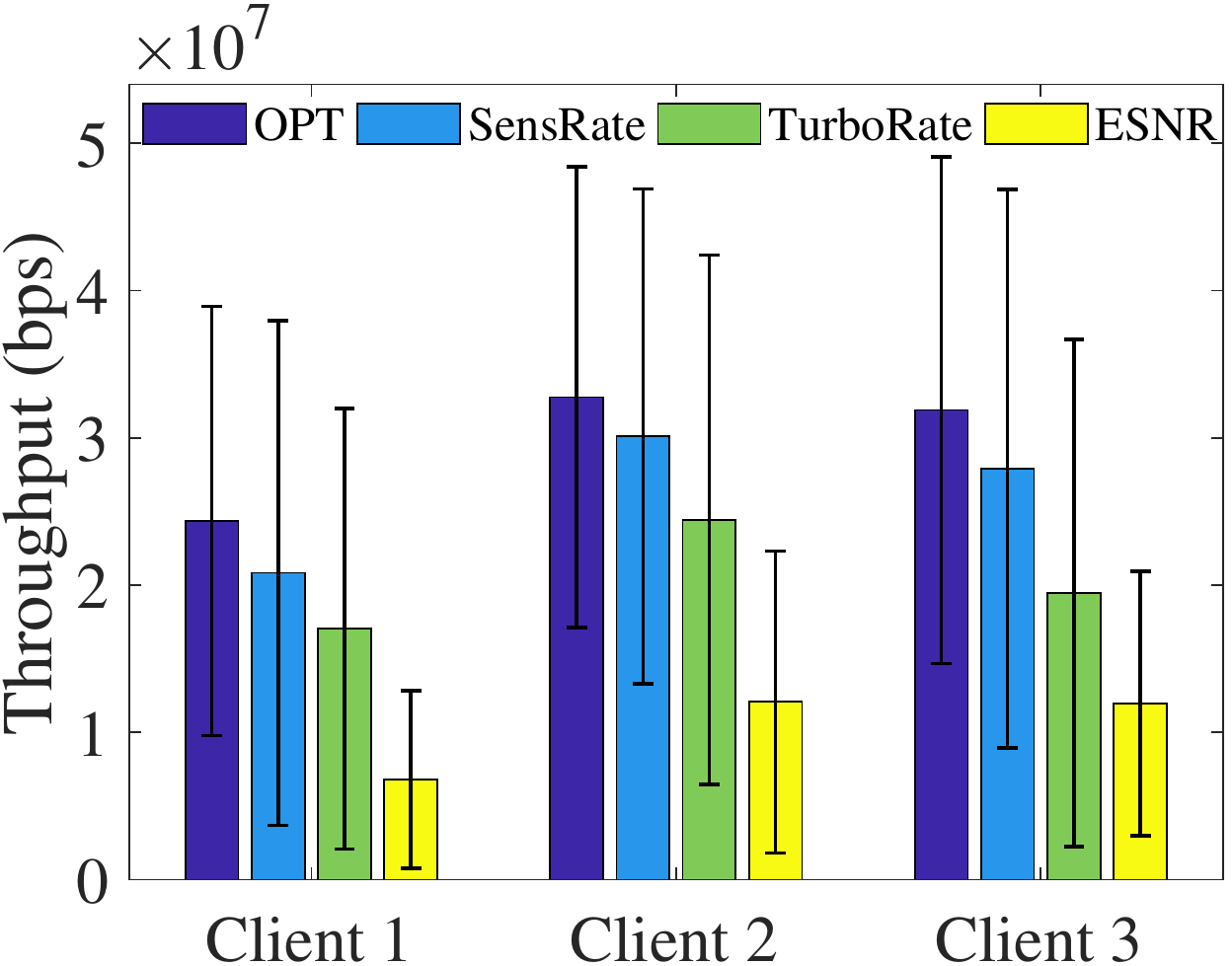}\label{3an-th}}
	\\
	\caption{The comparison between different RA algorithms for 2- and 3-antenna APs. }\vspace{-3mm}
	\label{} 
\end{figure}

\subsection{Overall System Throughput}

To begin with, we test the overall throughput of SensRate and compare it with several baseline algorithms defined in Section~\ref{sec:evaluation.a}. The whole experiments are conducted in 2-antenna AP and 3-antenna AP scenarios, respectively. 
 
\textbf{2-antenna AP scenario.} We first focus on the 2-antenna AP scenario, where two single-antenna clients concurrently transmit to a mobile UAV hotspot. The two clients are randomly deployed in the wild, with combination of dots in Fig.~\ref{trajectory_3d}. 
For each choice of clients' locations, we collect the real-time sensor data and CSI for over ten vertical and horizontal UAV trajectories, respectively. The UAV velocities vary from 0 to 10~m/s and the UAV-to-client distances are within 45~m. We use the CSI reading rate $f_r$ of 50~Hz to evaluate SensRate when the UAV velocity is below 6~m/s and increase $f_r$ to 100~Hz when the UAV velocity is over 6~m/s, which is also used in following experiments. 
We conduct the same experiment on both sites shown in Fig.~\ref{square} and Fig.~\ref{parking}.

The results in Fig.~\ref{2an-th} illustrate that SensRate can increase throughput for both clients. The overall throughput gain reaches 1.24$\times$ and 1.77$\times$ over TurboRate and ESNR, respectively. This gain mainly benefits from the SensRate's prediction function and adaptability to the UAV flights. With the assistance of sensor data, SensRate can 
provide a more accurate $\text{SNR}_{proj}$ prediction value albeit with the CSI staleness.

\textbf{3-antenna AP scenario.} Next, we check the system performance in the 3-antenna AP scenario, where three single-antenna clients concurrently transmit to the mobile UAV.
We repeat this experiment with the same configurations as in the 2-antenna AP scenario.
The throughput of four systems is demonstrated in Fig.~\ref{3an-th}. SensRate delivers an overall throughput gain of 1.28$\times$ and 2.48$\times$ over TurboRate and ESNR, respectively. 
Consistent with the 2-antenna AP scenario, the capability of SensRate to adapt to dynamic UAV channels allows all clients to choose more accurate transmission rates and achieve higher throughput.

\begin{figure*}[htb]
	\centering
	\subfigure{
		\includegraphics[width=0.235\linewidth]{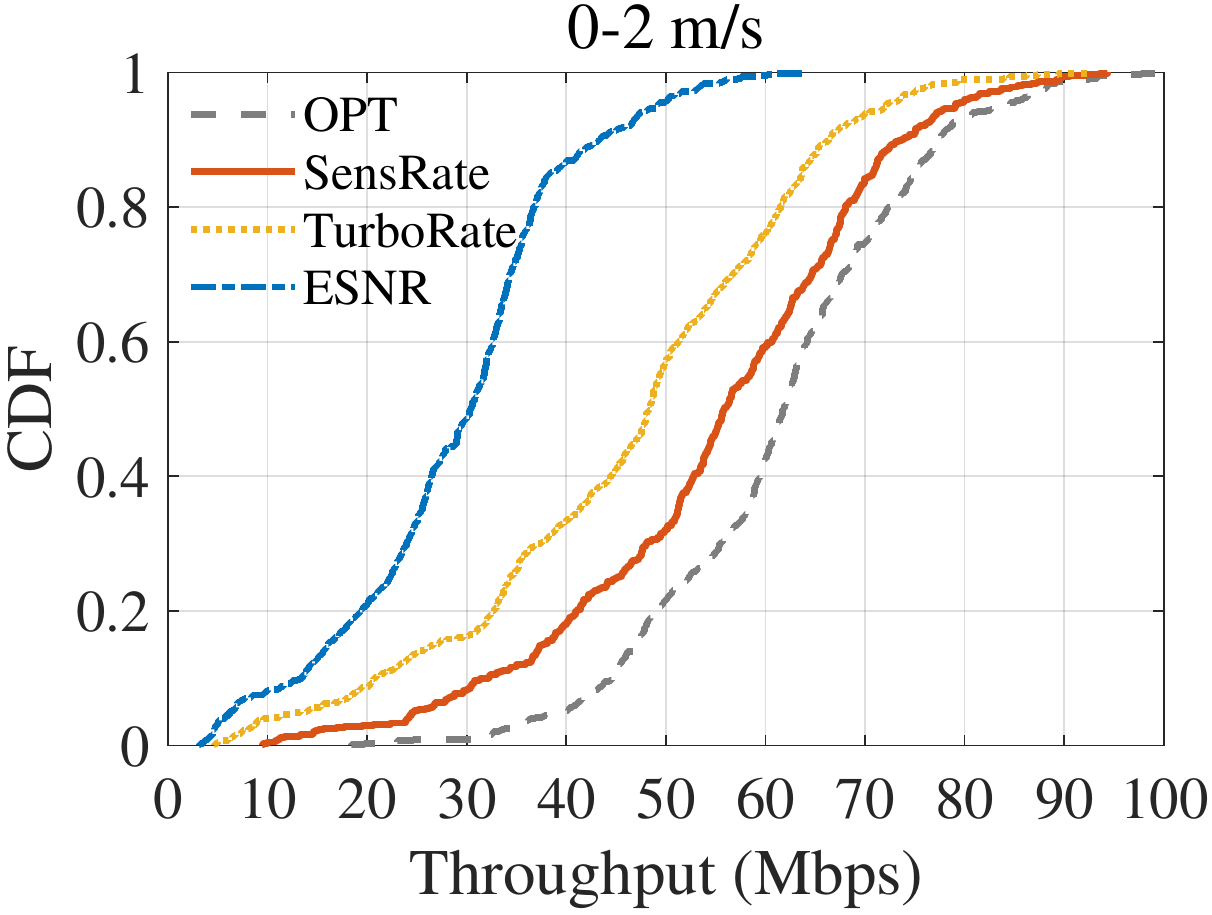}\label{th-vel1}}
	\hfill
	\subfigure{
		\includegraphics[width=0.235\linewidth]{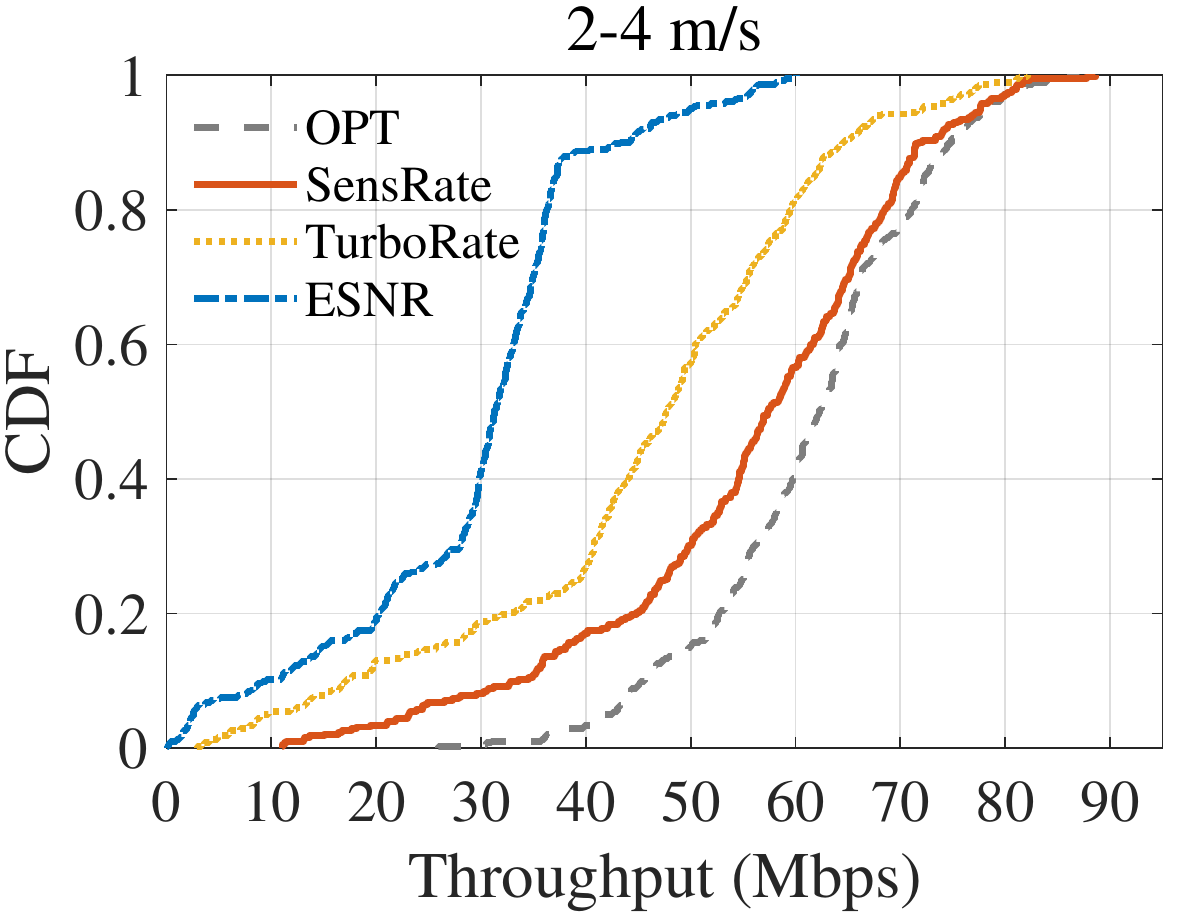}\label{th-vel2}}
	\hfill
	\subfigure{
		\includegraphics[width=0.235\linewidth]{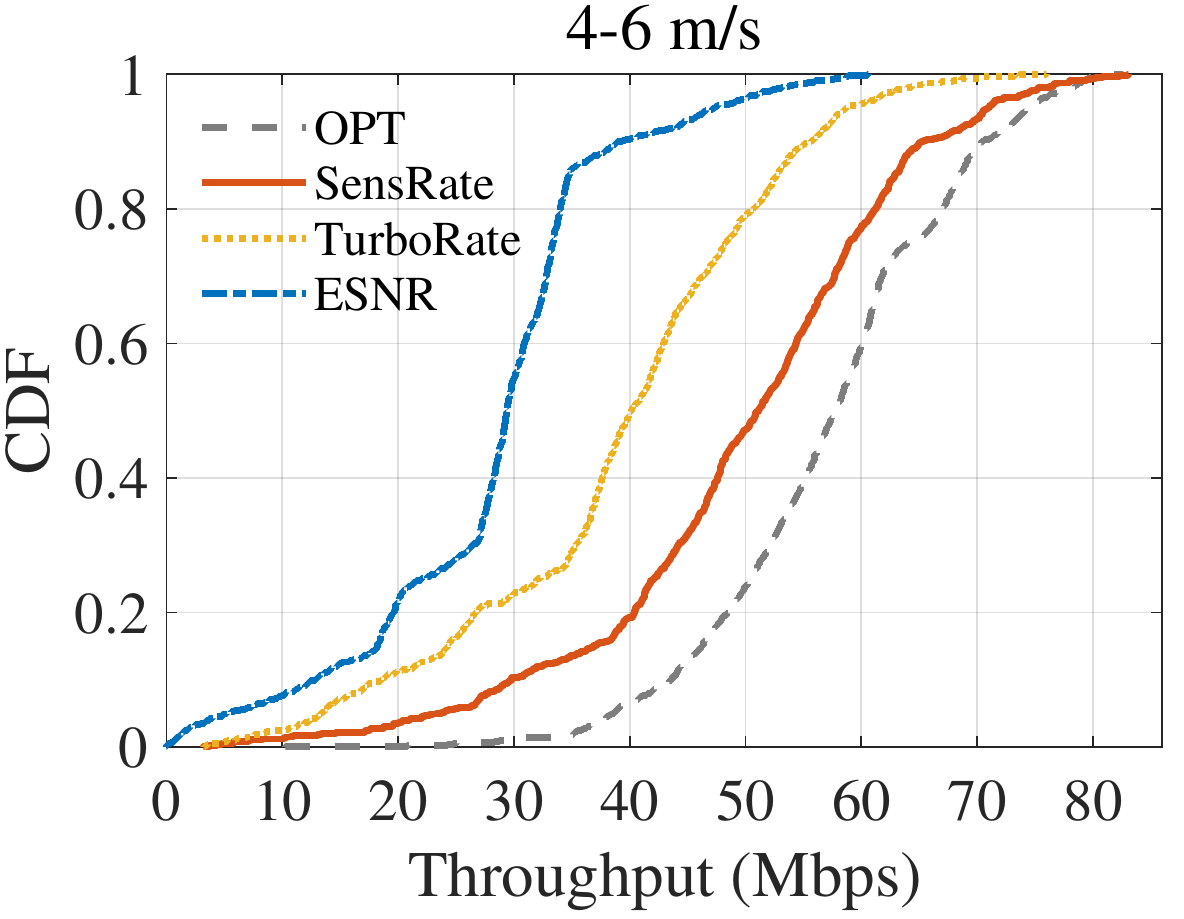}\label{th_vel3}}
	\hfill
	\subfigure{
		\includegraphics[width=0.235\linewidth]{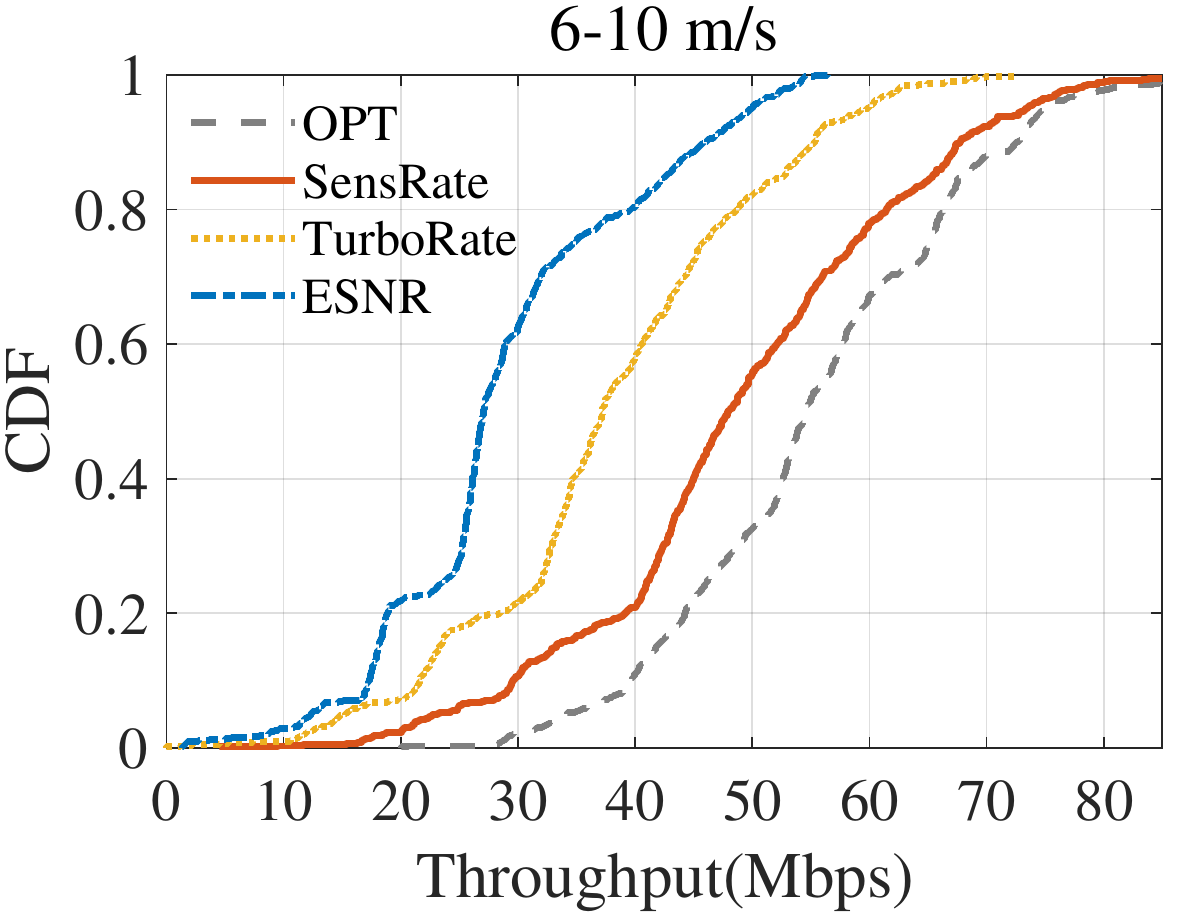}\label{th_vel4}}
	\\
	\vspace{-0.2cm}\caption{The comparison between different RA algorithms across varying UAV velocities.}
	\label{th-vel} 
\end{figure*}

\begin{figure*}[htb]
	\centering
	\vspace{-0.3cm}\subfigure{
		\includegraphics[width=0.235\linewidth]{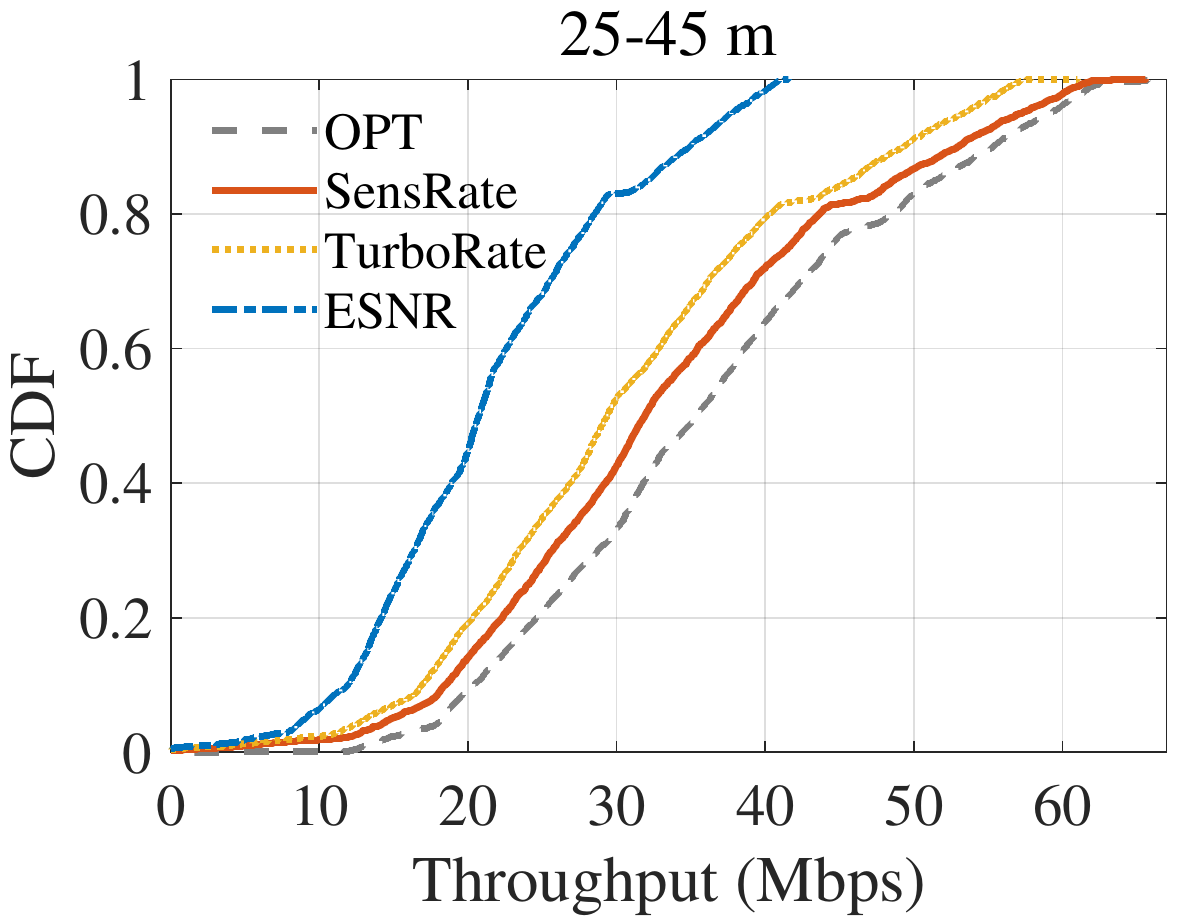}\label{th-dis4}}
	\hfill
	\subfigure{
		\includegraphics[width=0.235\linewidth]{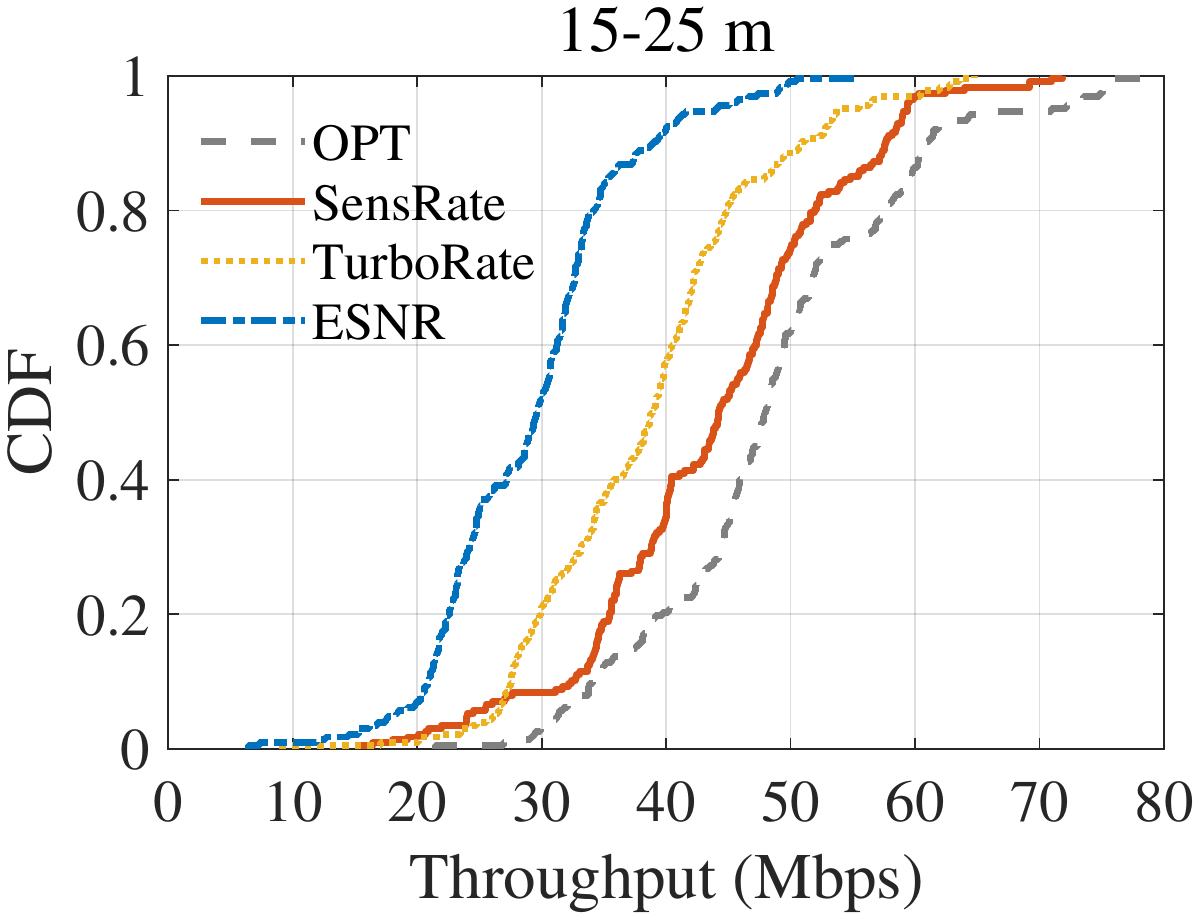}\label{th-dis3}}
	\hfill
	\subfigure{
		\includegraphics[width=0.235\linewidth]{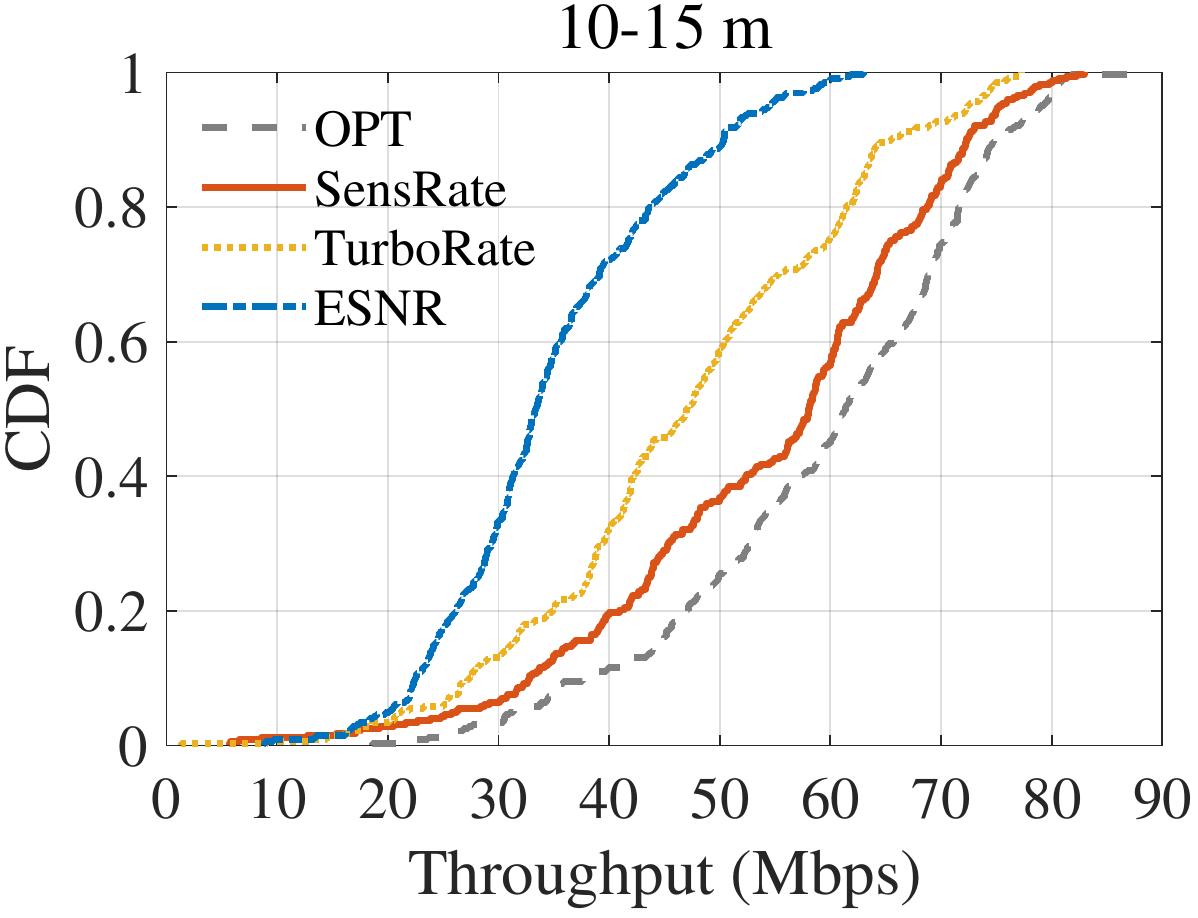}\label{th_dis2}}
	\hfill
	\subfigure{
		\includegraphics[width=0.235\linewidth]{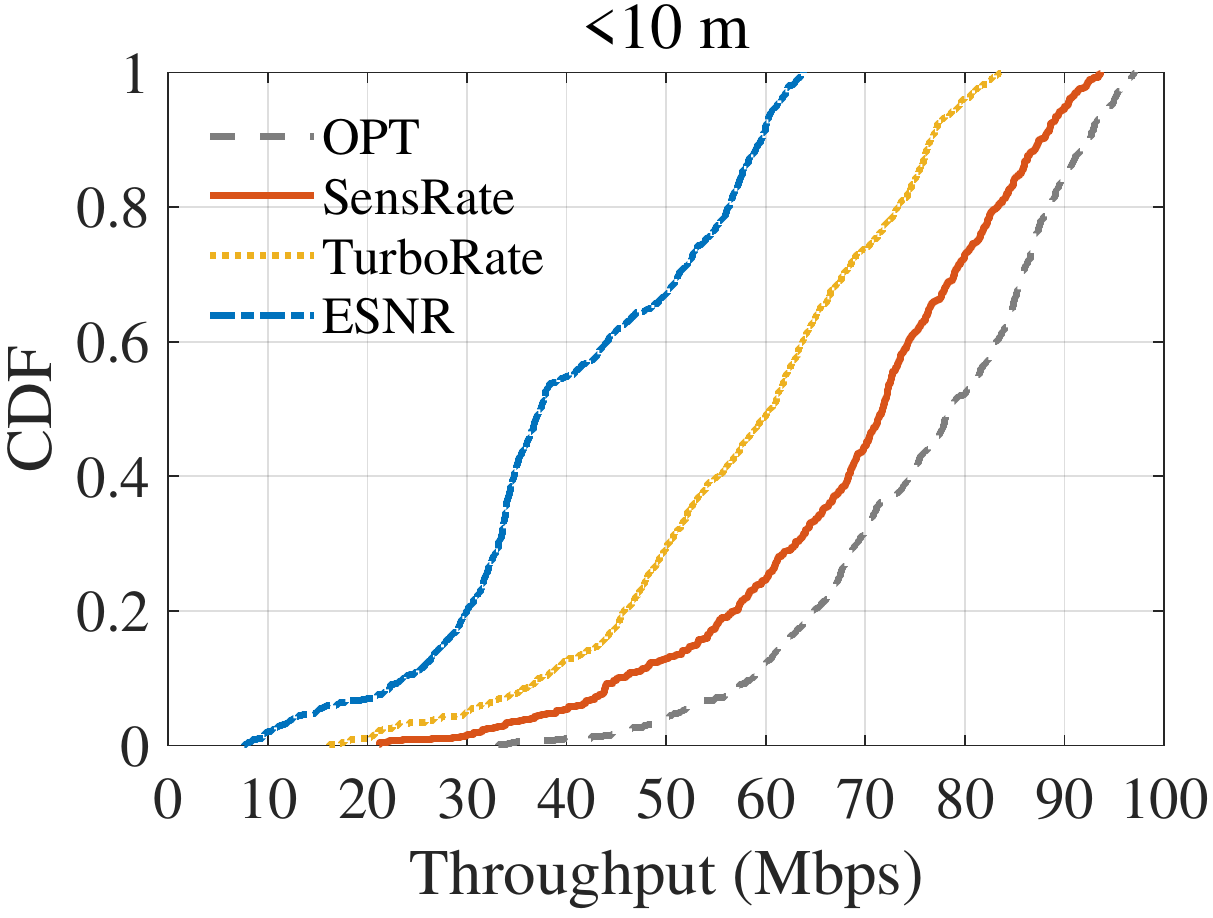}\label{th_dis1}}
	\\
	\vspace{-0.2cm}\caption{The comparison between different RA algorithms across varying UAV-to-clients distance ranges.}\label{th-dis}\vspace{-0.3cm}
\end{figure*}

\begin{figure}
	\centering
	\includegraphics[width=0.5\textwidth]{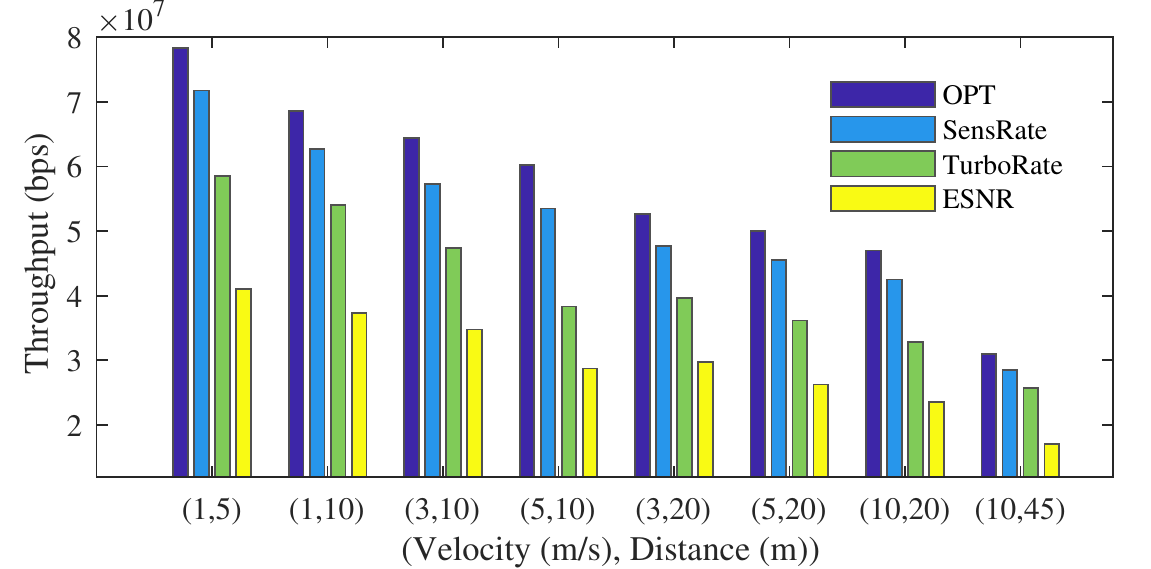}
	\vspace{-0.55cm}\caption{The results of changing distance and velocity comprehensively.}\label{th-vel-dis}\vspace{-3mm}
\end{figure}

\begin{figure*}[htb]
	\centering
	\subfigure[Trajectory 1]{
		\includegraphics[width=0.318\linewidth]{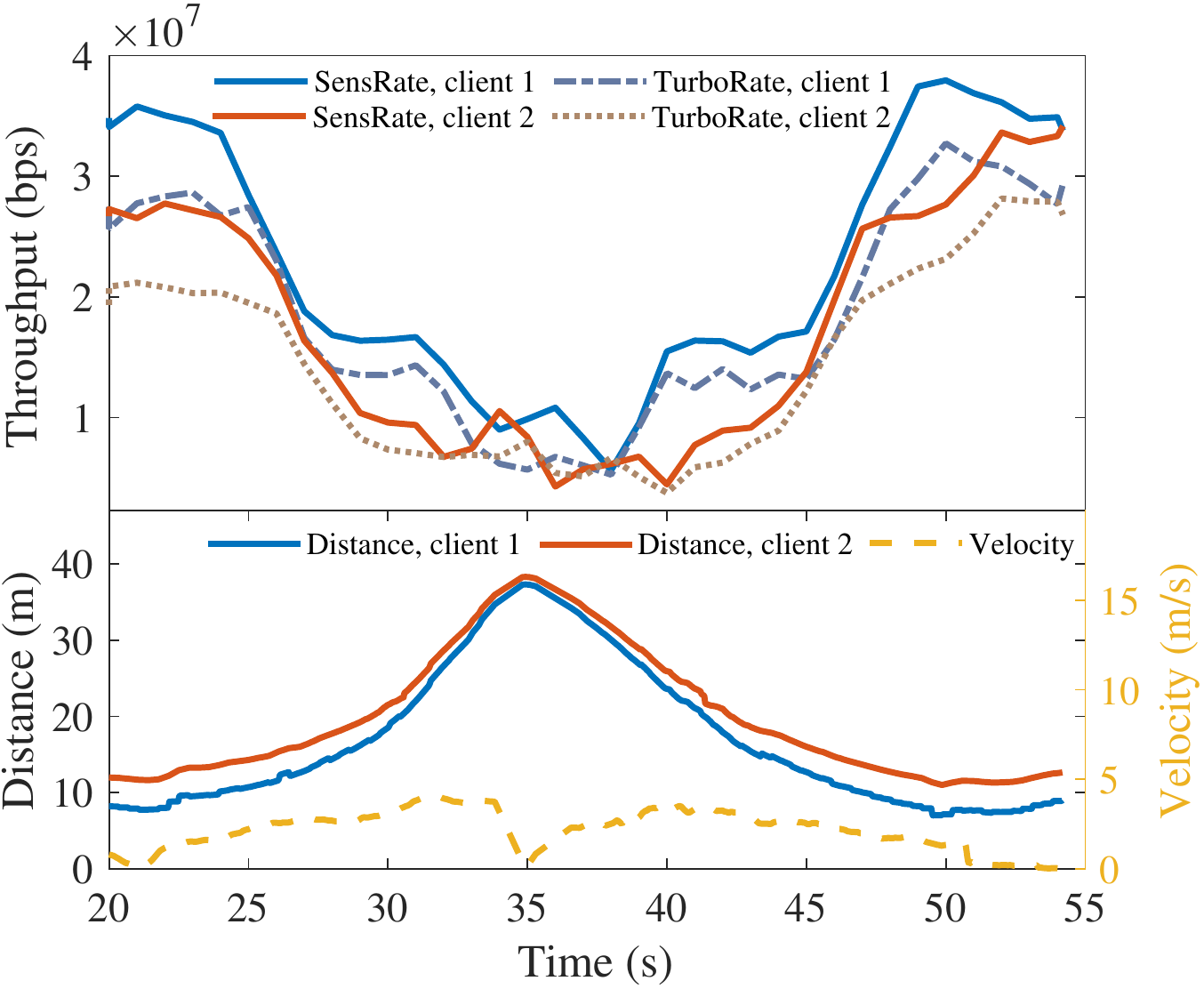}\label{trajectory1}}
	\hfill
	\subfigure[Trajectory 2]{
		\includegraphics[width=0.318\linewidth]{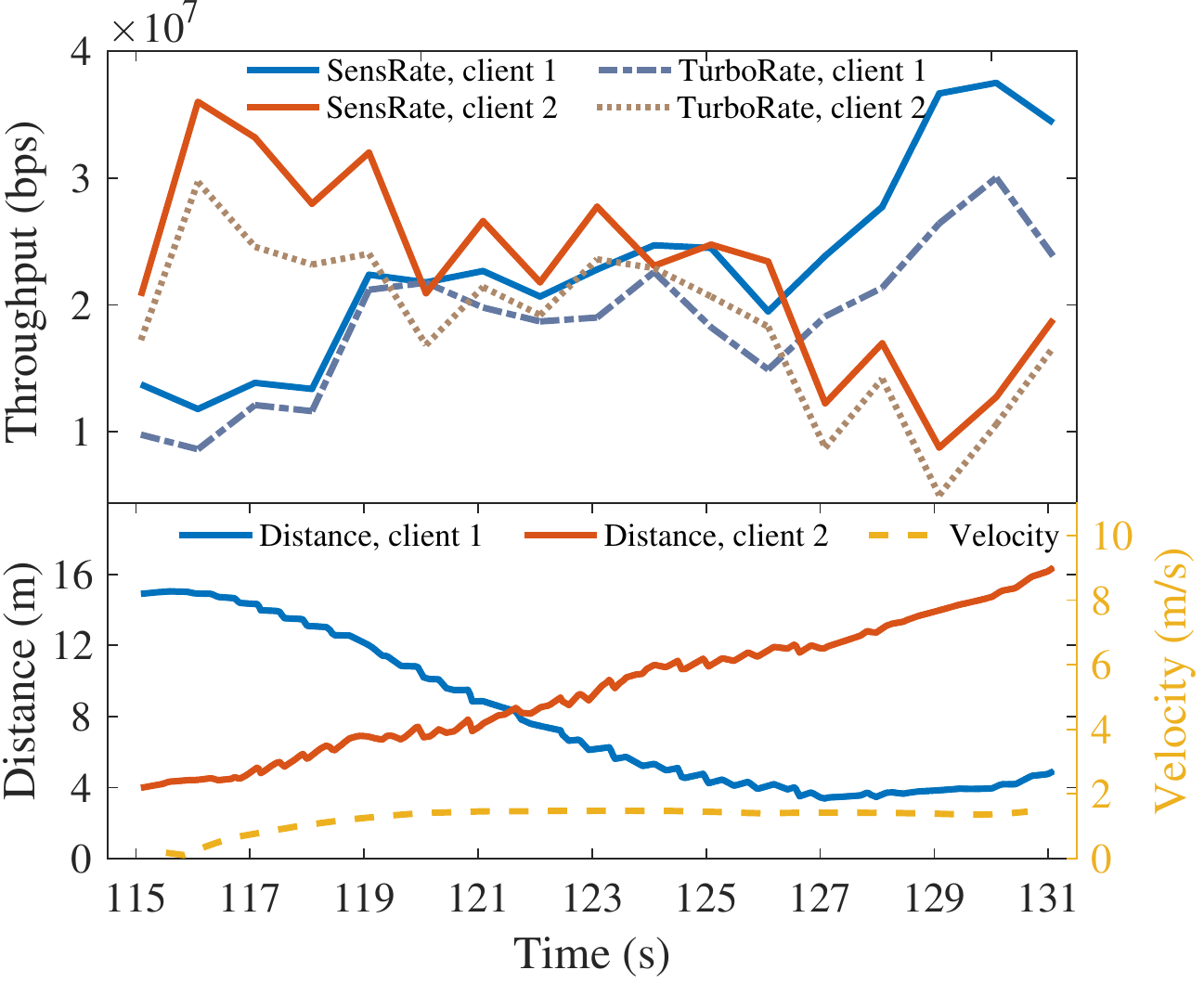}\label{trajectory2}}
	\hfill
	\subfigure[Trajectory 3]{
		\includegraphics[width=0.326\linewidth]{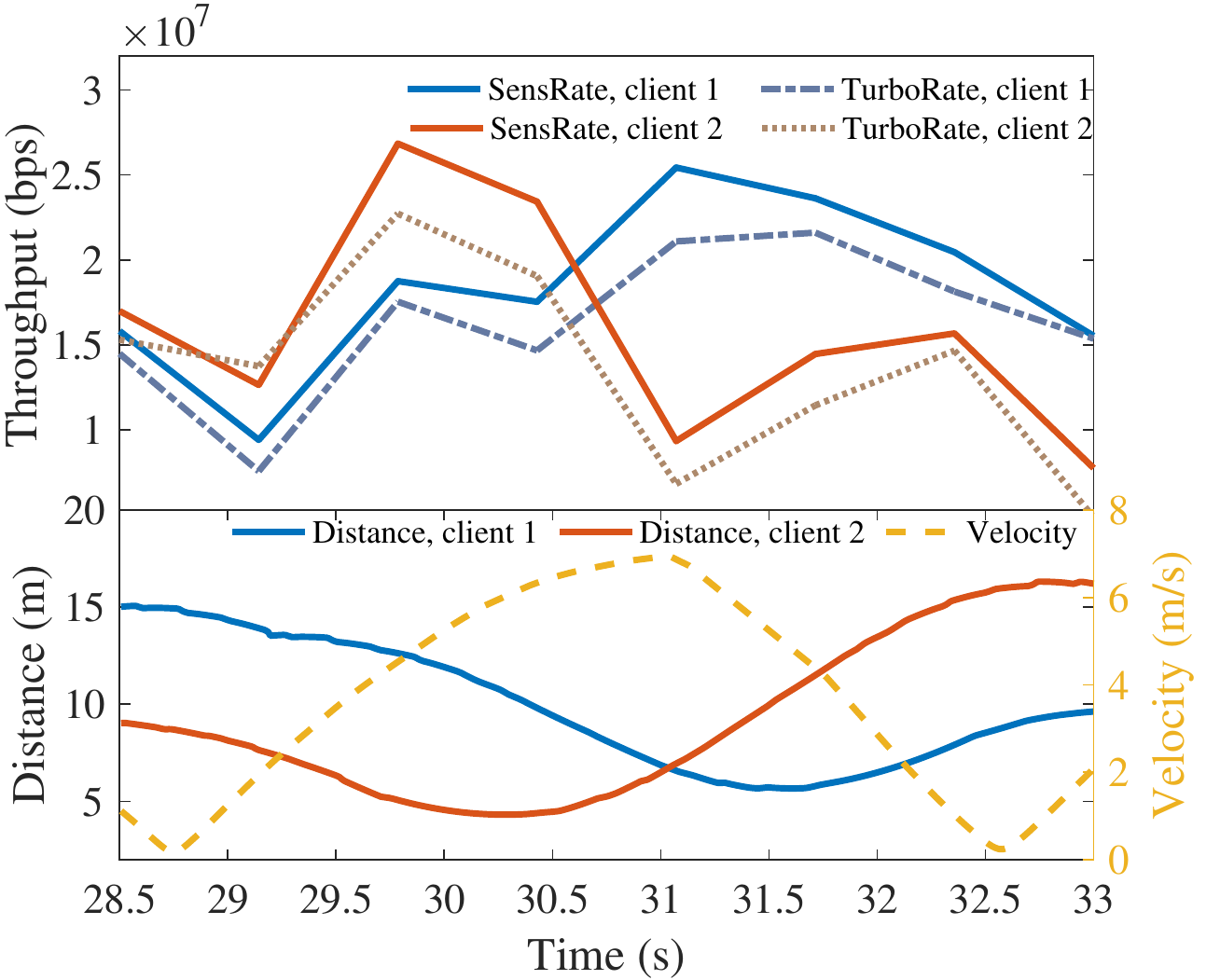}\label{trajectory3}}
	\\
	\vspace{-0.2cm}\caption{The dynamics of throughput and real-time flight states under different trajectories. (a) Trajectory 1. (b) Trajectory 2. (c) Trajectory 3.}
	\label{trajectory} \vspace{-0.3cm}
\end{figure*}

\subsection{Impact of Flight States}

We further compare SensRate with other RA algorithms in terms of 
different flight states. For simplicity, we zoom in on the overall system throughput in the 2-antenna AP scenario.

\textbf{Impact of velocity.} In this experiment, we locate the two clients at dots B and D in Fig.~\ref{trajectory_3d}, respectively, and control the UAV to move back and forth on trajectory 2 at different velocities of 0-10~m/s.
Fig.~\ref{th-vel} shows that SensRate outperforms TurboRate at each velocity range. The performance gap gradually increases with velocities, and reaches 26.7\% at UAV velocities of 6-10~m/s. Taking OPT as a reference, SensRate is able to maintain the overall throughput close to OPT, verifying the robustness in channel prediction function.

\textbf{Impact of distance.} We next test the performance of SensRate at different average distances between the UAV and two ground clients. We place two clients at dots A and B in Fig.~\ref{trajectory_3d}, and extend trajectory 2 to larger distance ranges, including \textless10~m, 10-15~m and 15-25~m. Then, the UAV altitude is increased to test the distance range of 25-45~m. The velocity is within 1-3~m/s. 
SensRate outperforms TurboRate at each distance range. The performance gap reaches the peak in the closest distance range, as the closer UAV-to-client distance increases the fading frequency, thus exacerbating the channel variance. SensRate can  offer strong adaptability. However, the closest distance also inevitably increases the difficulty for SensRate to predict the channel changes as accurately as in the case when the channel is stable, which slightly enlarges the gap between SensRate and OPT.

Fig.\ref{th-vel-dis} evaluates the system performance by comprehensively changing UAV-to-client distances and velocities. 
The throughput level of all systems decreases with velocities and distances. SensRate outperforms other existing RA algorithms and the gap between SensRate and TurboRate is the largest at the state tuple of (5,10), which causes larger channel variance. However, the throughout gain at the state tuple of (10, 45) significantly decreases, as the received power will no longer be dominated by the fast fading when the distance increases (e.g., to 45~m). 
The dependence of RA algorithms on channel prediction is reduced.  
Besides, the performance of SensRate is close to the OPT algorithm. In detail, the overall throughput of SensRate across different flight states can achieve 88.1\% to 93.5\% of the OPT algorithm.

\textbf{Evaluation for different trajectories.}
The above experiments verify the overall adaptability of SensRate. Then we further explore the specific throughput changes of clients under different UAV trajectories (shown in Fig.~\ref{trajectory_3d}), which are 

\begin{itemize}
	\item 
	\textbf{Trajectory 1:} The UAV flies vertically back and forth between the altitude of 2~m to 38~m at 2-4~m/s. The clients 1 and 2 are located at dots D and C, respectively.
	\item \textbf{Trajectory 2:} The UAV flies from client 2 (dot B) to client 1 (dot D) horizontally at a constant velocity of 1.3~m/s.
	\item \textbf{Trajectory 3:} The UAV passes by between client 1 (dot D) and client 2 (dot C) at variable velocities.
\end{itemize}

Fig.~\ref{trajectory} illustrates the dynamics of throughput results and real-time flight states. We have the following observations:

\begin{itemize}
	\item 
	In Fig.~\ref{trajectory1}, the throughput of both clients decreases simultaneously as the UAV's altitude increases. SensRate achieves throughput gains for each client throughout the flight compared with TurboRate, and the gains are most significant when the UAV's altitude is low at the beginning and end of the trajectory.
	\item In Fig.~\ref{trajectory2}, the throughput changes of clients 1 and 2 are opposite, with one increasing and the other decreasing as the UAV flies from one to the other. SensRate also outperforms TurboRate throughout the flight, whereas the peaks of the gains for clients are not synchronized, both depending on the time when the UAV is closest. 
	\item  Fig.~\ref{trajectory3} shows the result of a more complex UAV flight with changes in both distance and velocity. As expected, 
	the performance for each client is improved the most when the UAV is fastest and very close to the client, which causes severe channel fluctuations.
	
\end{itemize}

\subsection{Impact of CSI Reading Rates}\label{channel_prediction}

SensRate predicts MU-MIMO channel based on past CSI measurements. It is obvious that sufficient CSI readings can make the channel prediction mechanism perform better. However, to enable more CSI available at clients requires more UAV broadcast packets, 
which may hamper the channel utilization. In this subsection, we downsample the collected CSI traces to different rates of CSI readings $f_r$, including 200~Hz, 100~Hz, 50~Hz, 25~Hz and 10~Hz. Then, we investigate their impact on the system throughput and explore the minimum rate that meets SensRate prediction requirement.

\begin{figure}[t]
	\centering
	\includegraphics[width=0.5\textwidth]{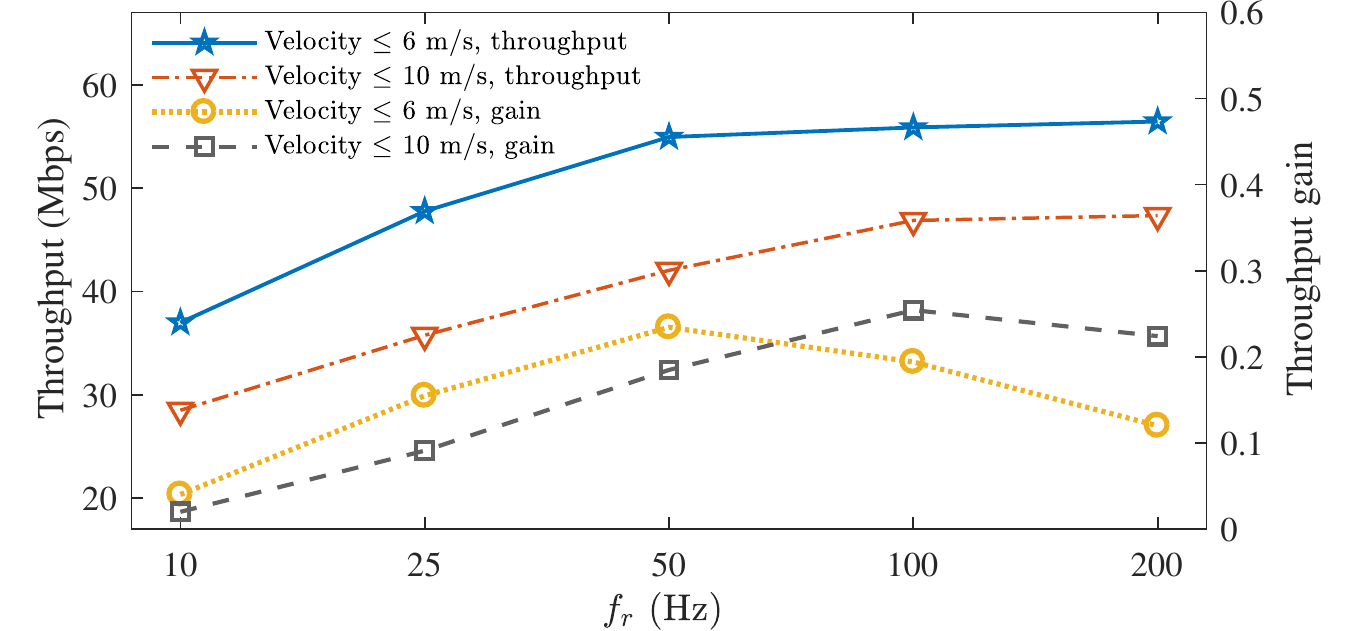}\vspace{-2mm}
	\caption{The impact of CSI reading rates.}\label{th-rate}\vspace{-3mm}
\end{figure}

Fig.~\ref{th-rate} compares the performance of SensRate and TurboRate under different $f_r$. 
When the maximum UAV velocity is 6~m/s, both the throughput and throughout gain increase rapidly with $f_r$ from 10~Hz to 50~Hz. When $f_r$ reaches 100 Hz or upper, the growth of the overall throughput significantly
slows down, and the gain gradually diminishes. 
These results indicate that 50~Hz CSI reading rate is sufficient for channel prediction in SensRate when UAV velocities are within 6~m/s. Continuing to increase the CSI reading rate may cause oversaturation and
threaten channel utilization. When the UAV velocity reaches 10~m/s, the overall throughput still significantly increases with $f_r$ from 50~Hz to 100~Hz, and then becomes relatively stable when $f_r$ is higher than 100~Hz.
Thus, we increase $f_r$ to 100~Hz to support SensRate at 6-10~m/s, as enough sample points between subsequent deep fades is required for the prediction algorithm to perform well.

\subsection{Impact of Environments}
In this subsection, we investigate whether SensRate adapts to environments with more  multipath 
reflections. We compare $\text{SNR}_{proj}$ prediction accuracy on both sites shown in Fig.~\ref{square} and Fig.~\ref{parking}, and test cases when the UAV flies at 1~m/s and 5~m/s on average. 
Table~\ref{table1} shows that whether the UAV is at 1~m/s or 5~m/s, the $\text{SNR}_{proj}$ prediction accuracy is nearly
the same on both square and parking lot.
This verifies that SensRate allows operation in the presence of various multipath reflections and adapts to different flight states.

We further analyze the tolerance of SensRate over environmental complexity by simulating the prediction error. The prediction error is partly quantified by the deviation of the SNR curve from the expected second-order polynomial based on the two-ray ground propagation model. As $P_r^{N} \propto (\frac{\lambda}{4\pi d_d})^2 \left|1+\sum_{i = 0}^{N-1}\rho_i \frac{1}{\gamma_i}exp(\frac{-j*2\pi(\gamma_i-1)*d_d}{\lambda}) \right|^2$, where $N$ is the number of reflected paths, $i = 0$ the reflected path from the ground, the SNR deviation is estimated by $D_N = 10*lg(P_r^{N})-10*lg(P_r^0)$. Since the SNR change exceeding 3~dB may require the node to adjust the rate~\cite{turborate}, we use 3~dB as the threshold. When $\text{E}[\left|D_N\right|] < 3~\text{dB}$, this environmental complexity can be highly tolerated. 
As it is impossible to simulate all states of reflectors, we simply assume the change of each $\gamma_i$ as $\gamma_i = k_i\gamma_0$. 
When $k_i>1$, through traversing all values of $k_i$, we summarize in Table~\ref{table2} the states of reflectors in a complex environment which SensRate can highly tolerate. When $k_i \leq 1$, i.e., the reflector $i$ is closer to the client than the ground, the approximate frequency $\gamma_0-1$ of the reflected wave from the ground is higher than that of other reflected waves, and the amplitude $\frac{1}{\gamma_0}$ and $\frac{1}{\gamma_i}$ is nearly the same. Thus, the occurrence of fading is generally consistent with our prediction. When there are not many reflectors with $k_i<1$, SensRate can highly tolerate this case.

For environments where the reflectors do not meet the above requirements, the tolerance of SensRate may be lower. However, the prediction in SensRate is realized by fitting CSI readings and the parameters are continuously updated. Thus, even if more complex environments may cause the fading pattern to deviate from the expectations based on the two-ray ground propagation model, the prediction function will still correct itself to reduce the error. 
Compared to traditional algorithms that directly use previously measured CSI to select rate, SensRate shows more promising for throughput gain.

\subsection{Moving Clients}
In order to verify the compatibility of SensRate to the moving clients, we further conduct two sets of experiments, in which the clients are stationary or being held by persons who walk randomly. As the ground clients in our experiment (NUC) do not have IMU or GPS, we employ the wireless-assisted technique in \cite{10.1145/3411833} to estimate velocity $v'$ in the direct-path direction and UAV-to-ground distance $d_d$. This process is detailed in Section~\ref{sec:design.c}. We can see the results in Table~\ref{table1}. When the UAV flies at 5~m/s, the moving clients have little impact on the throughput and prediction error. When the UAV flies at 1~m/s, the impact increases slightly, as the random walk of clients increases the non-linearity degree of the relative trajectory between the UAV and ground clients, which increases the difficulty of channel prediction. However, this performance degradation is not much, and there is still an obvious throughput gain compared to TurboRate, especially at UAV velocity of 5~m/s. This reflects the adaptability and robustness to moving clients and non-linear flight trajectories.

\section{Related Work}\label{sec:related}
Related works fall into the following categories.

\textbf{MU-MIMO RA algorithm.}
Most existing MU-MIMO RA algorithms rely on  
per-user PER~\cite{sur2016practical,6990336} or CSI ~\cite{turborate,xie2013adaptive,Scalable,lin2017acpad}. TurboRate~\cite{turborate}  
measures uplink CSI through periodic broadcast from the AP to select uplink rates. 
Some works~\cite{xie2013adaptive,Scalable,lin2017acpad} assign the AP to receive CSI feedback from clients to determine downlink rates and user selections. They~\cite{xie2013adaptive,Scalable} focus on finding ways to reduce CSI overhead and improve channel utilization. Some studies~\cite{sur2016practical,6990336} adjust downlink rates according to per-user PER that updated upon the reception of ACK frames. Such PER-based RA algorithms eliminate CSI feedback overhead, which have been favored by large 802.11 vendors such as Qualcomm and Broadcom for simplicity. 
All these RA algorithms rely on the channel measurements in coherence time without prediction, which are not optimized for the dramatically fluctuating UAV channels.

\begin{table}[t]
	\renewcommand\arraystretch{1.4}
	\centering
	\caption{Performance of SensRate under Different Environments and Moving Clients.} \label{table1}
	\vspace{-1.5 mm}
	\setlength{\tabcolsep}{0.2mm}{
		\begin{tabular}{c|c||c|c|c|c}
			\hline  
			\makecell[c]{Environment, UAV \\ Velocity (m/s)}&\makecell[c]{Prediction \\ error}&\makecell[c]{Client , UAV \\ velocity (m/s)}&\makecell[c]{Prediction \\ error}&\makecell[c]{Throughput \\ (Mbps)}&\makecell[c]{Gain}\\
			\hline %
			Square, 1& 0.5212& Stationary, 1 & 0.5133 & 58.0386 & 20.23\%\\ 
			Parking lot, 1& 0.6523 & Moving, 1 & 0.6429 & 55.8218 & 15.6\%\\ 
			Square, 5& 1.9085 & Stationary, 5 & 1.9079 & 50.2342 & 27.4\%\\
			Parking lot, 5&2.0092 &  Moving, 5 & 1.9310  & 49.8839 & 26.5\%\\
			
			\hline
	\end{tabular}}\vspace{-2mm}
\end{table}

\textbf{Inter-user interference prediction.} The prediction, adaptation and cancellation of inter-user interference play important roles in MU-MIMO networks. 
For UL streams, TurboRate~\cite{turborate} exchanges channel directions to estimate inter-user interference and adjust rates.
Zhou et al.~\cite{signpost} prioritize clients whose channel directions are aligned with predefined orthogonality vectors to minimize inter-user interference. MIMOMate~\cite{mimomate} selects clients with minimum interference among them. For downlink streams, many studies~\cite{lin2017acpad,8413159} enhance the channel utilization by optimizing precoders to suppress inter-user interference.
However, these estimations are still calculated by past channel measurements, which have been experimentally verified in~\cite{signpost} to be easily affected by mobile devices. 

With the rapid development of deep learning (DL), researchers have advocated applying DL in MIMO channel prediction~\cite{zhu2019adaptive,yuan2020machine,luo2018channel,han2020deep}. Some works~\cite{zhu2019adaptive,yuan2020machine} predict by exploiting temporal correlation across CSI series. OCEAN~\cite{luo2018channel} leverages the spatiotemporal relationship of CSI and considers frequency, location, etc. However, the distance is divided into sub-regions. YOLO~\cite{han2020deep} performs DL-based MIMO channel reconstruction by viewing the channel as an image. Whereas, none of them dive into the fine-grained correlation between motion states and MU-MIMO channel changes. Using DL to achieve sensor-assisted MU-MIMO channel prediction can be our future work. SensRate lays the foundation for it.

\textbf{Aerial channel prediction.}
Several studies~\cite{6863654,8048502,7835273} model the path loss and shadowing effect to determine optimal UAV placements. 
Some works~\cite{8125764,8432501} predict the received power based on the distance between aircraft pairs. 
DroneFi~\cite{chowdhery2018aerial} predicts the SNR by capturing the periodic fading patterns along flights, based on the two-ray propagation model. DroneNet~\cite{dronenet} combines the 3-D ray tracing and throughput. He et al.~\cite{staterate} exploit the UAV sensor data to train a neural network for channel prediction and link adaptation. However, it focuses on the single-user transmission and cannot give a fine-grained correlation to support channel prediction at any transmission time. In contrast, SensRate builds a model with time as the axis and sensor data as the parameter to predict aerial channel changes.
Recent efforts~\cite{312809,rajashekar2018beamforming} measure the delay spread of air-to-ground multipath channels for different UAV altitudes and elevation angles. Willink et al.~\cite{7079507} characterize the air-to-ground MIMO channels and examine the spatial correlation across antenna arrays, which take into account the phase but do not form a prediction method. 
To summarize, few researchers pay attention to the prediction of channel directions or phases of UAV MIMO networks, making it difficult to estimate inter-user interference and adapt rates.

\begin{table}[t]
	\renewcommand\arraystretch{1.22}
	\centering
	\caption{States of Reflectors in A Complex Environment Which SensRate Can Highly Tolerate.} \label{table2}
	\vspace{-1.5mm}
	\setlength{\tabcolsep}{3mm}{
		\begin{tabular}{c|c}
			\hline  
			\makecell[c]{$N$}&\makecell[c]{Requirement of $\gamma_i,~i=1,2,3,...,N-1$}\\
			\hline %
			$N = 2$ & $ \gamma_1 \geq 1.4\gamma_0$\\ 
			$N = 3$ & $ \gamma_1 \geq 1.5\gamma_0,~ \gamma_2 \geq 2.4\gamma_0$\\ 
			$N = 4$ & $\gamma_1 \geq 1.7\gamma_0,~ \gamma_2 \geq 2.3\gamma_0,~ \gamma_3 \geq 3.4\gamma_0$\\
			$N = 5$ &  $\gamma_1 \geq 1.8\gamma_0,~ \gamma_2 \geq 2.6 \gamma_0,~ \gamma_3 \geq 2.9 \gamma_0, ~ \gamma_4 \geq 3.7\gamma_0$\\	
			\textbf{ .~.~.}&\textbf{ .~.~.}\\	
			\hline
	\end{tabular}} \vspace{-2mm}
\end{table}

\section{Discussion}\label{discussion}

\textbf{Underlying applications.}
SensRate is proposed for low-altitude UAVs. The underlying applications include traffic offloading in hotspot areas~\cite{chowdhery2018aerial,entrepreneur1,entrepreneur2,AT&T,bushnaq2020optimal,zhang2021energy,arzykulov2021uav}, data collection~\cite{samir2019uav,bushnaq2019aeronautical} and UAV-enabled mobile edge computing~\cite{zhang2018stochastic,zhou2018computation}, considering the UAV altitudes and velocities in this paper. For example, DroneFi~\cite{chowdhery2018aerial} proposes drone hotspots flying below 20~m. Tethered UAVs have recently attracted attention for network coverage~\cite{bushnaq2020optimal} with limited altitudes and velocities. For applications of data collection and mobile edge computing, the UAV altitudes set in related works generally vary from 10~m to 100~m~\cite{samir2019uav,bushnaq2019aeronautical,zhang2018stochastic,zhou2018computation}. Moreover, a maximum UAV velocity of 10~m/s is widely used in studies of these UAV applications~\cite{chowdhery2018aerial,zhang2018stochastic,zhou2018computation,bushnaq2020optimal,zhang2021energy}, which is sufficient for UAV trajectory scheduling.

\textbf{Rotation.} There is no special optimization of SensRate to adapt to UAV rotation, but the harm can be minimized. The reason is that seamless switching can be achieved between SensRate and the traditional mode (i.e., using past CSI for rate selections without prediction). Once a client detects obvious UAV rotation, excessive non-linearity in relative motion trajectories or a sharp increase in prediction errors, the client can switch to the traditional mode in time to minimize performance degradation. Furthermore, the study of UAV rotation can be our future work to further optimize UAV transmission.

\section{Conclusion}\label{conclusion}

This paper introduces SensRate, a sensor-assisted UL MU-MIMO RA algorithm tailored for mobile UAVs to maximize the overall throughput. SensRate mitigates the impact of CSI staleness and enables the channel direction and inter-user interference to be traced and predicted under agile UAV mobility. 
We think this is an important design point for UAV-MIMO to materialize the high throughput and enhance the performance of UAV hotspots.
The experimental results show that SensRate achieves an average throughput gain of 1.24$\times$ and 1.28$\times$
over the best-known RA algorithm for 2- and 3-antenna APs, respectively. 
We hope this design can contribute the MU-MIMO communication in mobile scenarios by providing new insights on channel prediction and link adaptation.

\balance
\bibliographystyle{unsrt}
\bibliographystyle{IEEEtran}
\bibliography{IEEEabrv,./uav-mimo}

\end{document}